\documentclass[iop, twocolappendix, appendixfloats]{emulateapj}
\usepackage{graphicx}
\usepackage{multirow}
\usepackage[fleqn]{amsmath}
\usepackage{natbib}
\usepackage{hyperref}
\hypersetup{
     colorlinks   = true,
     allcolors    = blue
}

\begin{document}

\title{Impact of Instrument Responses on the Detectability of One-point Statistics from Redshifted 21~cm Observations}
\author{Piyanat Kittiwisit, Judd D. Bowman, Daniel C. Jacobs, Nithyanandan Thyagarajan, Adam P. Beardsley}
\affil{School of Earth and Space Exploration, Arizona State University}
\email{piyanat.kittiwisit@asu.edu}

\begin{abstract}
We study the impact of instrumental systematics on one-point statistics (variance, skewness, and kurtosis) of redshifted 21~cm intensity fluctuation observations from the Epoch of Reionization. We simulate realistic 21~cm observations based on the Murchison Widefield Array (MWA)  Phase I reionization experiment, using the array's point spread function (PSF) and antenna beam patterns, full-sky 21~cm models, and the FHD imaging pipeline. We measure the observed redshift evolution of pixel probability density functions (PDF) and one-point statistics from the simulated maps, comparing them to the measurements derived from simpler simulations that represent the instrument PSFs with Gaussian kernels. We find that both methods yield one-point statistics with similar trends with greater than 80\% correlation for all statistics. We perform additional simulations based on the Hydrogen Epoch of Reionization Array (HERA), using Gaussian kernels as the instrument PSFs, and study the effect of frequency binning on the statistics. We find that PSF smoothing and sampling variance from measuring the statistics over limited field of view dilute intrinsic features and add fluctuations to the statistics but, at the same time, reveal new detectable features. Particularly, observed kurtosis will increase when a few extremely high or low temperature regions are present in the maps. Frequency binning reduces the thermal uncertainty but can also blur regions along the frequency dimension, resulting in kurtosis peaks that only appear in statistics derived from maps of certain frequency bins. We further find that the kurtosis peaks will reach their maxima when the angular resolution of the PSFs match the size scale of the extreme regions that produce the peaks. The HERA array should be capable of charting the evolution of the observed skewness and kurtosis of the 21~cm fluctuations with high sensitivity while the MWA Phase I will likely only be capable of detecting the peak in variance.
\end{abstract}

\section{Introduction}
\label{sec:intro}

Considerable effort is underway to constrain the Epoch of Reionization (EoR), the era when radiation from the first stars and galaxies transformed gas in the intergalctic medium (IGM) from neutral to ionized. Observations of the Lyman-$\alpha$ forest in high-redshift quasars have set a limit to the end of reionization of $z\sim6.5$ \citep{2006AJ....132..117F}, and measurements of the cosmic microwave background (CMB) optical depth by the Planck experiment have indicated that reionization is still progressing at $z\sim8.8$ \citep{2016A&A...594A..13P}.

The 21~cm emission from the hyperfine transition in the ground state of neutral hydrogen is arguably the most direct probe to detect the EoR \citep{1972A&A....20..189S,1990MNRAS.247..510S,1997ApJ...475..429M,2000ApJ...528..597T,2002ApJ...572L.123I}. Full-sky observations of 21~cm spectra, redshifted to meter-wave, will produce tomographic maps of neutral hydrogen throughout the reionization era and beyond, allowing the study of the evolution of this structure and its implication for the underlying ionizing sources.

Many telescopes have been built or are being built to conduct experiments hoping to map this signal; these include the MWA (Murchison Widefield Array; \citet{2013PASA...30....7T,2013PASA...30...31B}), LOFAR (Low Frequency Array), PAPER (Donald C. Backer Precision Array for Probing the Epoch of Reionization; \citet{2010AJ....139.1468P}), HERA (Hydrogen Epoch of Reionization Array; \citet{2016arXiv160607473D}), and the SKA (Square Kilometer Array; \citet{2013ExA....36..235M}).

These telescopes utilize many compact antennas to yield wide fields of view and point spread functions (PSF) with approximately arcminute angular resolutions. These characteristics are ideal for EoR tomographic mapping but also give rise to widefield beam chromaticity and strong sidelobe interferences that complicate mitigation of bright astrophysical foregrounds.

Thus, much attention has been focused on the statistical analysis of redshifted 21~cm EoR observations, in particular with power spectrum measurements. Preliminary results from current observations have recently been released \citep{2016arXiv160806281B,2016ApJ...818..139T,2015PhRvD..91l3011D,2015ApJ...809...61A,2015ApJ...809...62P,2015ApJ...801...51J}, including robust characterization of the foreground contamination and instrumental systematics in the power spectrum \citep{2015ApJ...807L..28T,2015ApJ...804...14T,2013ApJ...769....5J}. Lessons from the first generation of experiments have led to the design of HERA, which will be a highly-packed, redundant-baseline array that is optimized for the power spectrum analysis, while preserving imaging performance on sub-degree scales.

As reionization progresses, ionized regions (H\,{\scriptsize II}) will form around groups of sources with high-energy UV radiation, causing the distribution of 21~cm intensity field to deviate from Gaussian, and the power spectrum alone will be insufficient to fully describe the signal \citep{2007ApJ...659..865L}. This limitation has led to studies of several other statistics. 

One promising alternative is the one-point probability distribution function (PDF) and higher-order moments (variance and skewness) of the 21~cm brightness temperature fluctuations. Simulations have suggested strong evolution of the PDF throughout the reionization redshifts \citep{2004ApJ...613...16F}, and the non-Gaussianity in the distribution can be captured by measuring the skewness of the PDF \citep{2007MNRAS.379.1647W,2009MNRAS.393.1449H}. \citet{2010MNRAS.406.2521I} has also developed a maximum-likelihood method to directly measure the shape of the PDF. \citet{2010MNRAS.408.2373G} further used such a method to differentiate the PDF from six different reionization models. Recently, \citet{Watkinson:2014jv,Watkinson:2015ce} and \citet{2015MNRAS.449.3202W} studied the sensitivities of the variance and skewness on different reionization scenarios, including the global and local reionization models, sinks of reionization radiation, and Lyman $\alpha$ coupling and X-ray heating. They suggest that signatures in the statistics could be detected by upcoming instruments. Although not discussed here, other efforts to constrain non-Gaussianity in the 21~cm signal have explore more-complicated statistics, including the two-point difference probability density function (\citet{2008MNRAS.384.1069B}) and the 21~cm bispectrum (e.g. \citet{2006MNRAS.366..213S}, \citet{2007ApJ...662....1P}, \citet{2015MNRAS.451..266Y}, and \citet{Shimabukuro:2016ea} among others).

Yet characterization of foregrounds and instrumental systematics on the one-point statistics have not been well studied. The complicated PSFs of the reionization arrays are often ignored from the analysis or approximated by convolving the model images with ideal Gaussian PSF. Noise from foreground residuals and other systematics are usually added to the simulation as Gaussian random noise or analytically derived from theory. Real observations also pose other limitations such as the limited field of view, the effect of frequency bandwidth, and the evolution of the signal along the frequency dimension. These issues must be robustly studied before predictions can be made for the upcoming data.

In this paper, we study some of the above issues in detail by performing realistic simulations that closely mimic actual EoR observations, combining full-sky 21~cm models, readily available radio array simulators, and other software packages from the data reduction pipelines of the MWA and HERA. We focus our analysis and discussion on the detectability of the variance, skewness and kurtosis, taking into account the impact of instrument PSF, field of view, and frequency bandwidth. We describe our simulation method for the MWA in Section~\ref{sec:simulation}. We present our results and discuss the effect of the instrument PSF and field of view on the statistics in Section~\ref{sec:psf_effect}. In Section~\ref{sec:detectability}, we build HERA simulations based on results from previous sections and discuss the impact of frequency bandwidth on the statistics, as well as the potential of using kurtosis to detect ionized regions in 21~cm maps.

\section{Simulations}
\label{sec:simulation}

We construct our MWA simulation pipeline based on existing 21~cm models, a telescope simulator and the actual MWA imaging pipeline. We describe each component in the following subsections. All simulations are noiseless to simplify the interpretation of the results although mathematically derived uncertainties are included in the analysis. We also ignore foreground contamination, postponing it to a future work. Although we only discuss the MWA simulation here, our simulation pipeline can be adapted to different instruments.

\subsection{Full-Sky 21~cm Models}
\label{sec:21cm_model}

Neglecting foregrounds, the EoR sky consists of 21~cm brightness temperature intensity fluctuations ($\delta T_b(\nu)$), which are characterized by the 21~cm spin temperature ($T_S$),  the CMB temperature ($T_{\gamma}$), the fractional over-density of baryons $(1 +  \delta)$, and the gradient of the proper velocity along the line of sight ($\mathrm{d}v_{\parallel}/\mathrm{d}r_{\parallel}$),
\begin{equation}
    \begin{split}
        \delta T_b(\nu) \approx 9 x_{HI}&(1+\delta)(1+z)^{1/2} \\
        &\left[1 - \frac{T_{\gamma}}{T_S}\right]
         \left[\frac{H(z)/(1+z)}
                    {\mathrm{d}v_{\parallel}/\mathrm{d}r_{\parallel}}
          \right]\ \mathrm{mK}.
    \end{split}
\end{equation}

The current state-of-the-art 21~cm simulators such as the 21cmFAST use semi-analytic methods to produce three-dimensional cubes of 21~cm brightness temperature fluctuation at different redshifts \citep{2011MNRAS.411..955M}. These cubes are represented in rectangular comoving coordinates and can be up to a couple of (Gpc/h)$^3$ in size, roughly equivalent to $\sim10$~deg$^3$ volume at relevant redshifts. 

In contrast, EoR observations produce tree-dimensional data where two of the dimensions map the spatial dimensions of the sky in sine projected coordinates and one dimension conflates the telescopes' line-of-sight distances with time and redshift/frequency. The MWA observes the EoR from 138.915~MHz to 195.235~MHz with 80 kHz frequency channel resolution, equivalent to 704 maps that span $z\sim9.2-6.3$. Thus, the observed data form a single "lightcone" cube, where slices across the frequency dimension evolve with redshift.

In order to match existing 21~cm models to an instrumental observation, we transform the four-dimensional (three spatial and one redshift) outputs of theoretical 21~cm simulations into a three-dimensional (two angular and one frequency) 21~cm cube. We use a tile-and-grid method that maps the simulation cubes from different redshifts to full-sky maps in HEALPix\footnote{\url{http://healpix.jpl.nasa.gov}} coordinates with NSIDE=4096, capturing the full spatial structure and evolution of the signal in frequency-space.

We start with an original suite of ``raw'' 21~cm cubes from the semi-analytic simulations of \citet{2013ApJ...767...68M}. The cubes span the redshift range $13>z>6.2$, with $\Delta z=0.1$ steps, representing the universe from $2\%$ to $100\%$ ionized respectively. The simulation volume is $1\,\mathrm{Gpc}^3$ in a $128^3$ pixel box with periodic boundary. We linearly interpolate the simulated 21~cm cubes across redshift to produce new cubes that match the redshifts observed by the MWA at each of its frequency channels, yielding 704 cubes.  For each of the 704 redshifts, we tile the interpolated cube with itself to construct an arbitrarily large simulated volume for that redshift. Then, we draw an observable sky at that redshift as a sphere of radius equal to the comoving distance of that redshift from a fixed origin inside the volume and linearly interpolate the four nearest neighboring pixels from the cube to the corresponding HEALPix pixel location on the sphere. The HEALPix pixel area is $\sim10$ times smaller than the resolution of the simulations to avoid Nyquist sampling. This process is repeated for every observed redshift to produce a suit of full-sky maps that accurately represent the 21~cm lightcone model. The HEALPix maps are later re-projected into half-sky sine projected maps with pixel area $\sim1$ arcminute and passed to our observation simulator.

\subsection{Instrument Model}
\label{sec:instrument}

The main instrumental effects of a radio interferometer result from the array's PSF response which is both directional and spectral dependent. The size and shape of the PSF depend on the response of individual antenna elements (primary beam), the antenna layout, and the frequency and pointing of the observations. 

We model the PSF effects of the MWA Phase I \citep{2013PASA...30....7T}, which consists of 128 phased-array antenna tiles arranged over a region with a diameter of 3~km. Each tile is a four-by-four grid of dual-polarization dipoles. Nearly half of the antenna tiles are contained in a compact core with a diameter of 100~meters. EoR observations only use this compact core to improve EoR power spectrum sensitivity, foreground subtraction and calibration \citep{2013MNRAS.429L...5B,2009ApJ...695..183B}. This yields naturally-weighted PSF resolution $\sim1$ degree across the MWA EoR band. The primary beam extends across the whole sky, but the usable main lobe is about 30 degrees diameter centered around the pointing center.

The MWA acquires EoR data in a drift-and-shift mode. The telescope is pointed at a specific coordinate in a radio quiet region of the sky (for example, $\alpha=0$h, $\delta=-30^{\circ}$) and observes as the sky drifts throughs the primary beam. Then the telescope is repointed, and the process repeats. We adopt a simpler single-point observation where the telescope is always pointed at the zenith for our simulation. We model the limiting case of a single snapshot image.  This results in the worst-case PSF for the MWA with no rotation synthesis to improve UV coverage. 

We use the MIT Array Performance Simulator (MAPS\footnote{\url{http://www.haystack.mit.edu/ast/arrays/maps/}}) to simulate MWA visibilities.  MAPS generates beam responses from user-configurable antenna configurations, array configurations and observing parameters, and performs convolution with sky models. We configure MAPS to use the MWA 128-tile array layout, using a short cross-dipole on an infinite ground plane as an antenna model. MAPS also applies w-projection to correct for widefield effects. We run MAPS with our full-sky models for each observing frequency to take into account spectral dependency of the PSF. Data is integrated for two seconds.

The visibilities are then re-gridded and Fourier transformed by FFT into ``dirty'' maps of brightness temperature fluctuation with Fast Holographic Deconvolution (FHD) \citep{2012ApJ...759...17S}, one of the two primary calibration and imaging pipelines of the MWA \citep{2016ApJ...825..114J}. We assume that the derived dirty maps from FHD are the final data products that are used for EoR analysis. This is consistent with data from real observations in an absence of foregrounds. In principle, a deconvolution through a matrix inversion or an iterative algorithm such as CLEAN can be utilized to remove the PSF structure and recover the “true” intensity map. However, the sky structures in maps from reionization arrays are too complicated and not sufficiently sparse to utilize such an approach. For current EoR experiments, foreground-subtracted dirty images are generally where derived statistics are calculated.

\subsection{Gaussian Approximation of PSFs}
\label{sec:psf}

\begin{figure}
    \hspace{-3mm}
    \includegraphics[width=0.5\textwidth]{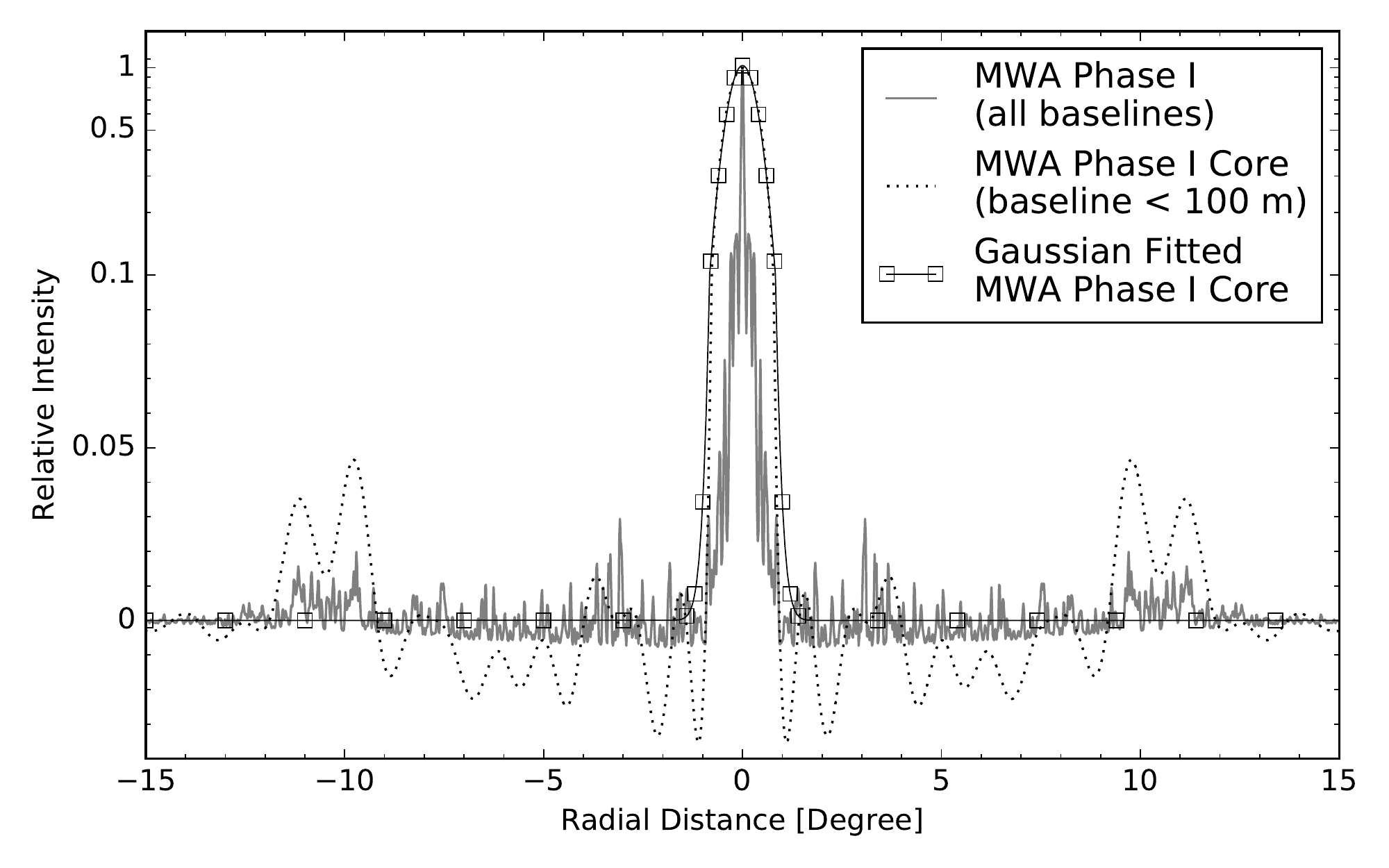}
    \caption{Cross sections of the beam responses of the full MWA Phase I with all baselines (solid line), and the MWA Phase I Core with only baselines less than 100 meters (dotted line). A Gaussian fit to the latter case is shown as solid line with squares. Note that scaling on the y-axis is linear from -0.05 to 0.1 and logarithmic otherwise.}
    \label{fig:psf}
\end{figure}

The use of idealized Gaussian models to represent telescope PSFs is common in previous studies. Gaussian kernels with FWHM equivalent to the angular resolution of the telescopes' PSFs are assumed as the telescope PSFs and convolved with sky models to produce ``observed'' intensity maps. This method neglects sidelobe responses and directional dependency of the PSFs. Figure \ref{fig:psf} shows these differences for the MWA. The grayed solid line shows a cross section of the PSF of the full MWA Phase I array that includes all antennas. The dotted line shows the PSF of the compact core that only includes antennas within the 100~meter diameter, which is the PSF used in EoR anlysis and in this work. The solid-line with squares shows a Gaussian fit to the main lobe of the compact core PSF that would be used in a simple convolution method. To investigate the differences between the more realistic and idealized PSFs on our analysis, we carry out simulations with both the MWA Phase I Core PSF, using FHD and MAPS, and its fitted Gaussian PSF, using a simple convolution. We fit a Gaussian kernel to the main lobe of the MWA Phase I Core PSFs derived from the beam configuration in FHD. Then, we convolve our two-dimensional Gaussian kernel to the sine projected 21~cm maps that we input to the MAPS-FHD pipeline. This produces similar output maps to the pipeline but without the the effects of the full instrumental PSF sidelobes..

\subsection{One-Point Statistics}
\label{sec:1point_stats_definition}

Our quantities of interest are the one-point statistics of brightness temperature fluctuations in the dirty maps, in particular variance, skewness and kurtosis. For a map with data value $x_i$ and mean value $\overline{x}$, the p-th order statistical moments are defined as,
\begin{gather}
    m_p = \frac{1}{N_{pix}}\sum_{i=0}^{N_{pix}} (x_i - \overline{x})^p.
\end{gather}
The variance ($S_2$), skewness ($S_3$) and kurtosis\footnote{In statistics, the precise term of this definition is excess kurtosis, which subtracts 3 from the standard definition of kurtosis, $m_4 / (m_2)^2$, to yield zero for Gaussian distribution. Here, we simply use kurtosis to refer to excess kurtosis.} ($S_4$) are standardizations of the 2nd, 3rd and 4th moments,
\begin{gather}
    S_2 = m_2, \\
    S_3 = m_3 / (m_2)^{3/2}, \\
    S_4 = m_4 / (m_2)^2 - 3.
\end{gather}

Along with the mean, these three quantities describe simple deviations in the shape of the brightness temperature PDF relative to a standard Gaussian PDF. The variance measures the spread of the PDF. Both skewness and kurtosis describe the outliers, or the tails, of the PDF in different ways. Skewness measures asymmetry of the outliers, in which positive and negative skewness values indicate that the values of the outliers are greater and less than the mean value that would be the case for a Gaussian distribution. Kurtosis describes the heaviness, or density, of the outliers; a positive kurtosis indicates more outliers whereas a negative kurtosis indicate less outliers. Thus, a PDF with high kurtosis will look more peaked in comparison to a Gaussian PDF with the same variance. A perfect Gaussian PDF has zero skewness and kurtosis.

We measure the PDF, variance, skewness and kurtosis from both the FHD and Gaussian simulated dirty maps, as well as from the residual maps, defined as the difference between the FHD and Gaussian maps. In real measurements, signal-to-noise ratio decreases substantially near the edge of the field due to instrument response. Therefore, we only use pixels within the FWHM of the primary beam to construct the PDFs and statistics. All PDFs are calculated using 60 linearly spaced bins from the minimum to maximum pixel values within each map. 

It is common in radio astronomy to oversample the PSF resolution and produce dirty maps with higher angular resolution than the PSF's angular resolution. The oversampled pixels contain no extra information as features smaller than the size of the PSF are filtered out from PSF smoothing. Mathematically, the oversampling of the PSF has no effect on the statistics measured in this work. Thus, we use all pixels in the dirty maps that fall within the FWHM to calculate one-point statistics. In Section~\ref{sec:detectability}, we will calculate thermal noise uncertainties, and in those calculations, we take care to use the proper number of independent PSF samples in the maps.

\begin{figure}
    \hspace{-4mm}
    \includegraphics[width=0.5\textwidth]{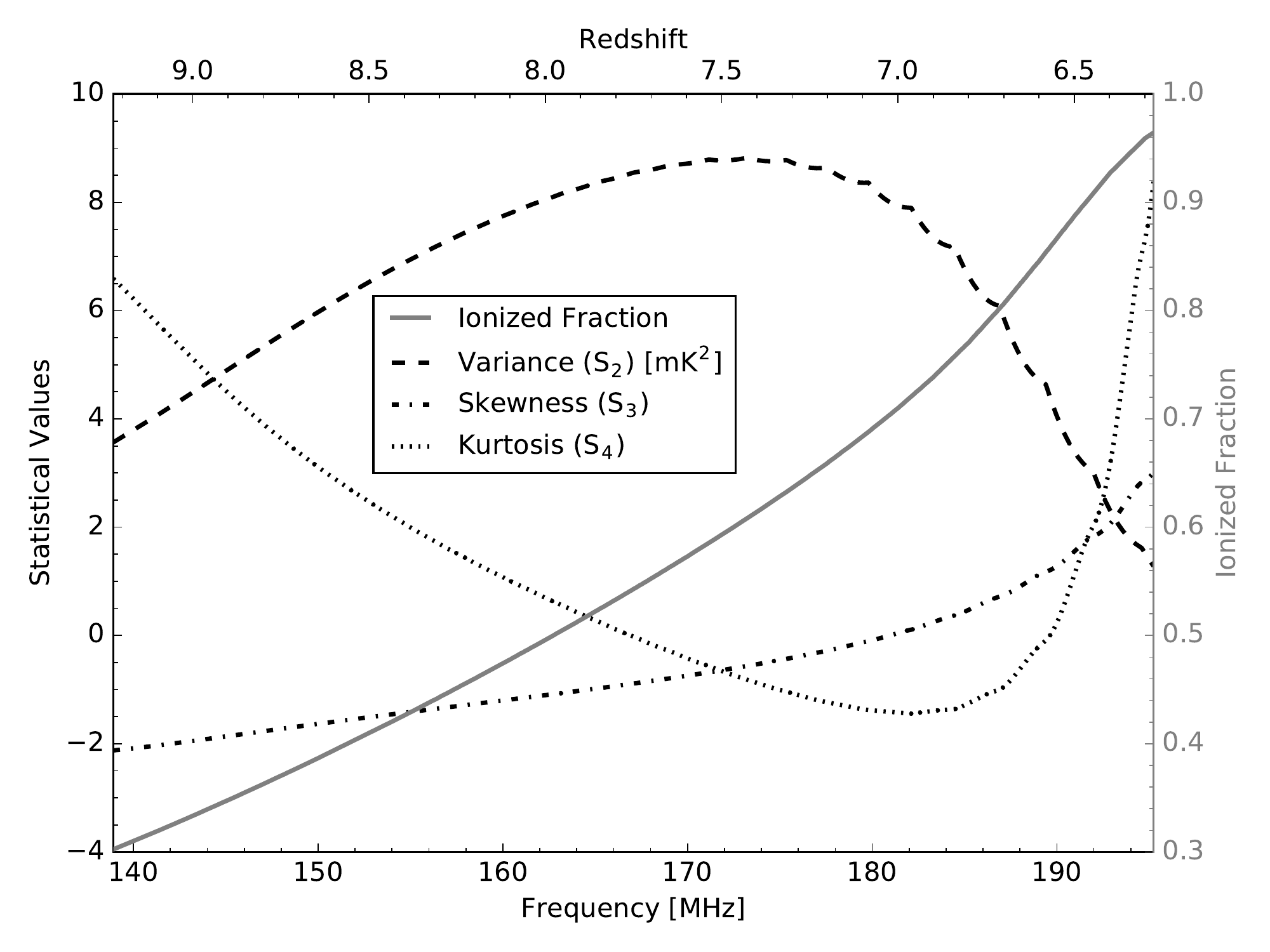}
    \caption{Redshift evolution of one-point statistics from the raw 21~cm input cubes. The dashed line, dash-dot line, and dotted line show variance, skewness and kurtosis respectively. Statistical values are shown on the left y-axis. The solid gray line shows the global ionization fraction as a function of redshift for the input 21~cm model and uses the right axis.}
    \label{fig:model}
\end{figure}

As a reference, we plot the variance (dashed line), skewness (dot-dash line) and kurtosis (dotted line) calculated from the raw 21~cm input model using all pixels, along with the ionized fraction of the cubes (solid line), as a function of frequency and redshift in Figure~\ref{fig:model}. The left y-axis shows corresponding statistical values, whereas the right y-axis shows the ionized fraction. Although not shown here, we find that statistics derived from the HEALPix and sine projected outputs of our grid-and-tile extrapolation closely match statistics from the raw model. The average fractional errors between statistics derived from the raw model and the HEALPix maps are less than 0.1\% with Pearson Correlation coefficients (PCC) greater than 0.99 (99\% correlated) for all statistics. Statistics derived from the sine projected maps diverge a bit more from the raw model with average fractional errors of 12\%, 0.8\% and 3.7\% for variance, skewness and kurtosis but still with PCC greater than 0.99. Specifically, the variance and kurtosis decrease due to interpolation effects. We will discuss and compare features in the model statistics with statistics derived from our simulations in Section~\ref{sec:1point_stats_features}.

\section{Effects of PSF}
\label{sec:psf_effect}

\subsection{Observable Features in One-Point Statistics}
\label{sec:1point_stats_features} 

\begin{figure}
    \hspace{-3mm}
    \includegraphics[width=0.5\textwidth]{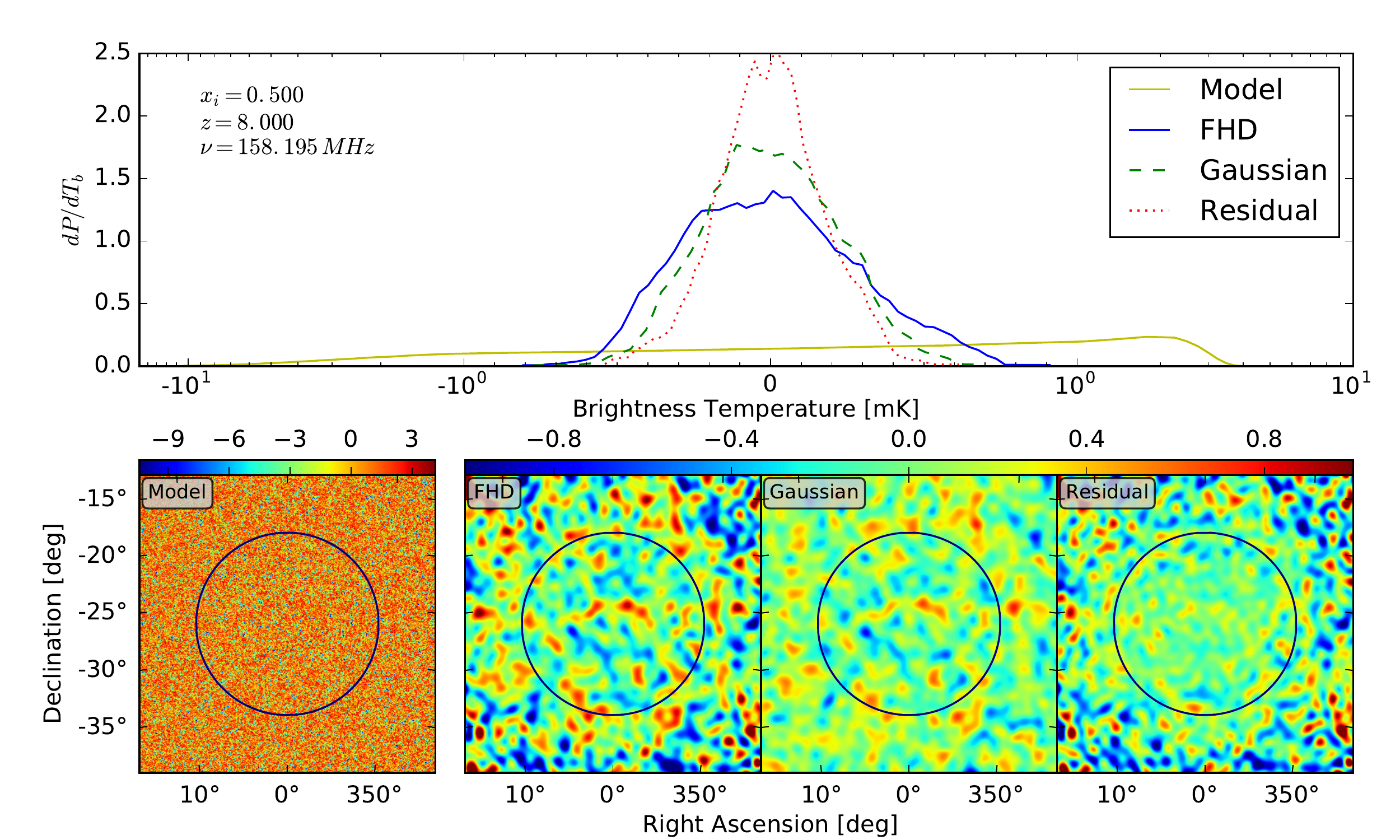}
    \caption{Example PDFs (upper panel) derived from the FHD (solid line) and Gaussian simulation (dashed line) at $x_i=0.5$ ($z=7.9$) along with the corresponding simulated maps (lower panels).  PDF of the residual map between the two simulations (FHD - Gaussian) is also shown as the dotted line along with the map. Black circles in the maps indicate the field of view of the telescope, equivalent to the FWHM of the primary antenna beam. Only pixels inside the field of view are used to calculate statistics.}
    \label{fig:mwa_pdf}
\end{figure}

Figure~\ref{fig:mwa_pdf} shows example PDFs derived from the FHD and Gaussian simulations at $x_i=0.5$ ($z=7.9$) along with the corresponding simulated maps. Two effects are readily apparent.  First, the lack of zero-spacing baselines in the MWA forces the observed sky brightness mean to equal zero.  As a result, all PDFs derived from simulated observations are offset from the PDFs of the input theoretical 21~cm model.  Second, the simulated observation PDFs are also more Gaussian-like than the input model.  Since there is no thermal noise in our simulations, this effect arises solely from the PSF smoothing that reduces the amplitudes of the intrinsic 21~cm fluctuations and blurs the relatively sharp transitions from ionized to neutral IGM in the input model as seen in the simulated maps.

\begin{figure}
    \includegraphics[width=0.48\textwidth]{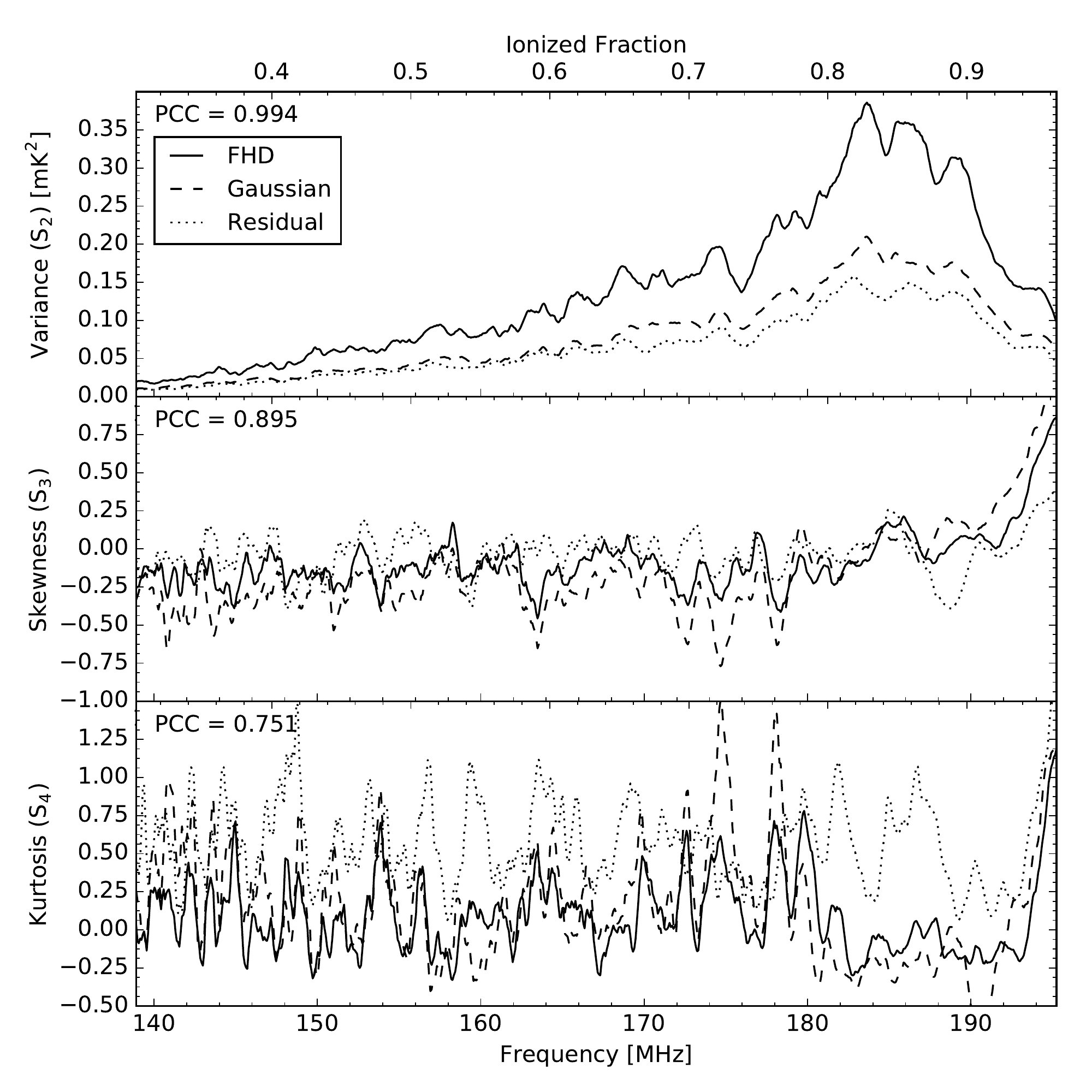}
    \caption{Variance (top), skewness (middle) and kurtosis (bottom) derived from the FHD (solid lines) and Gaussian (dash lines) simulations as a function of frequency and ionized fraction. Statistics for the residuals of the two simulations are also shown as dotted lines. PCC is the Pearson correlation coefficient between statistics derived from FHD and Gaussian simulations.}
    \label{fig:mwa_stats}
\end{figure}

Figure~\ref{fig:mwa_stats} shows the variance, skewness and kurtosis derived from the FHD (solid line) and Gaussian (dashed line) simulations as a function of ionization fraction (or frequency). The statistics derived from the simulated maps exhibit fluctuations throughout the measured redshifts, as opposed to smoothly varying measurements in the models, due to the sample variance inherent in simulating an observation of a single field, rather than modeling a statistical ensemble.


The variance is seen to steadily increase as reionization progresses, reaching a maximum around ionized fraction $x_i\approx0.85$, before falling off as reionization concludes. The intrinsic variance of the input model shows similar evolution, but reaches the highest point earlier, around $x_i\approx0.6$ (see Figure~\ref{fig:model}).   We interpret the shift in the variance between input and output as a result of the PSF smoothing.  The peak in the variance from the simulated observations does not occur until the ionized bubble size grows large enough to match the MWA Phase I Core's angular resolution.  This occurs later than the intrinsic peak in the input model.   We can relate this effect to power spectrum measurements.  The MWA Phase I Core only samples size-scales corresponding to low k-modes in Fourier space.  Hence, it is when these modes reach their peak power that determines when the variance peaks in the MWA observations.   The instrument is acting analogously to a matched-filter to identify signal features of a particular size scale.   In general, the redshift of the observed peak will be dependent on a telescope's angular resolution, and hence, may occur earlier (at higher redshift) for larger telescopes with higher angular resolution since ionized bubbles are smaller when the intrinsic variance in the signal peaks $x_i\approx0.6$. The PSF of the telescope also smooths out 21~cm fluctuations, lowering the signal amplitude. Thus, the observed variance peak has a lower amplitude than the intrinsic variance peak in the model.

The skewness derived from both simulated observations shows little evolution for most of reionization, remaining slightly negative around $-0.08$ and $-0.15$ mean values with standard deviation of $0.19$ and $0.30$ for FHD and Gaussian cases respectively.  At the end of reionization, however, it becomes increasingly positive until the last redshift in the simulation.  In contrast, the input model shows highly negative skewness early in the reionization due to the high filling factor of neutral hydrogen and rare regions of low emission, creating an offset positive peak in the PDF with a long tail extending toward zero emission. The model skewness becomes positive around $x_i\approx0.75$ when large ionized regions start to form and the PDF begins to become bi-modal.  The transition to positive skewness occurs later in the simulated observations, beginning at approximately the same time as the observed variance peaks, indicating that the skewness is diluted by the telescope's PSF smoothing until the ionized regions grow larger than the angular resolution.

The kurtosis also shows little evolution throughout reionization, oscillating around $0.09$ and $0.16$ mean with standard deviation of $0.25$ and $0.39$ for FHD and Gaussian cases, but does exhibit a few occasional spikes that rise above the standard deviation. For example, at $\sim175$ and $\sim178~MHz$. Upon examining the maps, we find that these spikes occur when the observed field contains a few large regions of extremely low or high brightness temperature (relative to the mean at that frequency). These outlier regions add more density to the tails of the PDFs, resulting in higher kurtosis.

Taking the near-zero values of the skewness and kurtosis together indicates that the observed distribution remains nearly Gaussian throughout most of reionization, in contrast to the evolving distribution intrinsic to the input model.  Hence, for the MWA Phase I, the power spectrum is a sufficient measurement to fully describe the information that is retrievable by the telescope from the underlying 21~cm signal, except at the end of reionization.

\subsection{Comparison of Gaussian and FHD PSFs}
\label{sec:psf_compare}

It is evident from Figures~\ref{fig:mwa_pdf} and~\ref{fig:mwa_stats} that the PDFs and one-point statistics resulting from the FHD-simulated maps and Gaussian-smoothed maps are qualitatively similar. Both PSFs produce similarly shaped PDFs. However, the FHD-derived PDF is consistently broader than the Gaussian-derived PDF and its variance is consistently higher by approximately $56\%$. We interpret the increased variance in the FHD case to be due to the complicated sidelobe structure inherent in the full instrumental PSF. This structure results in a sidelobe confusion effect that scatters power throughout the map, raising the overall variance and diluting the signal's skewness and kurtosis.

Despite the discrepency in variance between the two cases, the dependence on ionization fraction of each of the statistics is highly correlated between the two cases. We quantify the similarity by calculating the Pearson Correlation Coefficients (PCC) between the two cases for each one-point statistic. We find coefficients of $0.994$ for variance, $0.895$ for skewness, and $0.751$ for kurtosis.  

We attempt to further quantify the effect of sidelobe confusion by measuring statistics from the residual maps derived by subtracting FHD maps from Gaussian maps. The residual maps contain features from sidelobe confusion and other effects inherent in the full instrumental PSFs. Statistics derived from the residual maps are plotted in Figure~\ref{fig:mwa_stats} as dotted lines along with those from the FHD and Gaussian simulated maps. Variance of the residual maps closely follow variance from the Gaussian convolved maps as expected from the fact that the FHD-derive variance is 56\% higher than the Gaussian-derive variance. The residual skewness and kurtosis remains noise-like through the redshifts, oscillating around $-0.029$ mean and $0.136$ standard deviation for skewness and $0.576$ mean and $0.282$ standard deviation for kurtosis.

Based on these findings, we conclude that Gaussian-smoothing kernels are reasonable approximations to full instrument simulations for densely-packed antenna arrays such as the MWA Phase I Core. The skewness and kurtosis estimates derived from Gaussian smoothed maps are likely sufficient for many statistical investigations. For variance estimates, a simple transfer function of the form $m_2^{obs}=f(m_2^{Gaus})$ can be used to compensate for the factional difference between the estimates derived from realistic simulations and more-commonly modeled idealized cases.  Inspection of Figure~\ref{fig:mwa_stats} suggests that this function may reduce to a multiplicative correction factor such that $m_2^{obs}=f_{0} m_2^{Gaus}$, with $f_{0} = 1.56$ in our case. We have not explored different 21~cm input models, but given that most theoretical models yield similar features, we expect the correction factor to be generally independent of the detailed properties of the 21~cm signal for a given instrument.

\section{Detectability}
\label{sec:detectability}

Now we explore the detectability of the one-point statistics.  We will find below that the MWA Phase I does not have sufficient thermal sensitivity to detect skewness and kurtosis. Hence, building on our findings above that idealized Gaussian beam simulations are sufficient for dense aperture arrays, we perform new simulations based on HERA using only idealized Gaussian beams to focus our investigation on the detectability of the one-point statistics by a more-sensitive array.  We begin in Section~\ref{sec:hera} with an overview of the HERA telescope and our simulations of it.  We will then review our thermal uncertainty estimates in Section~\ref{sec:uncertainty} and discuss the effect of frequency bandwidth in Section~\ref{sec:bandwidth} before presenting our results.

\subsection{HERA}
\label{sec:hera}

HERA is a second-generation radio interferometer optimized for redshifted 21~cm power spectrum detection.  Presently under construction, HERA builds on the lessons-learned from the MWA and PAPER.  It consists of zenith-pointed, 14-meter diameter dishes fed by dual-polarization dipoles that are closely packed into a hexagon shape. The telescope has a $\sim$10-degree primary field of view and will operate between 100 and 200 MHz.  Nineteen dishes have been constructed to date as part of a staged build-out that will culminate in 331 dishes. \citet{2016arXiv160607473D} provides additional details about the design of HERA.

We model the HERA instrument response using its FWHM field-of-view and an idealized Gaussian smoothing kernel, corresponding to the angular resolution of the array, to approximate its PSF. We perform the Gaussian simulation with the final planned 331-dish HERA array, as well as a 37-dish HERA array that will be operable by 2017. The latter happens to match the angular resolution of the MWA Phase I Core, allowing for a relatively direct comparison of the two.  Table~\ref{tab:sim} summarizes our simulation parameters for the MWA Phase I Core and both HERA configurations.

\begin{table}
\begin{center}
\begin{tabular}{l c c c }
\hline
\multirow{2}{*}{Simulation Parameters} &MWA  &\multirow{2}{*}{HERA37} &\multirow{2}{*}{HERA331} \\
&Phase I Core & & \\
\hline
Number of antennas       &50     &37     &331        \\
Effective area (m$^2$)   &21.5   &153.86 &153.86     \\
Maximum baseline (m)     &100    &42     &140        \\
Integration time (h)     &1000   &100    &100        \\
Angular Resolution (deg) &$\sim$1-1.6
                         &$\sim$1-1.6
                         &$\sim$0.3-0.5 \\
FoV diameter (deg) &$\sim$20-35
                         &$\sim$7-11
                         &$\sim$7-11  \\
\hline
\end{tabular}
\caption{\label{tab:sim}Instrument specifications used in our simulations.  In addition to the parameters above, antenna layouts were taken from \citet{2012MNRAS.425.1781B} and \cite{2016arXiv160607473D}. For all of our simulations, we use a simulation bandwidth of $~\sim$139-195~MHz with 80~kHz raw spectral channel width.}
\end{center}
\end{table}

\subsection{Thermal Uncertainty}
\label{sec:uncertainty}

To estimate the impact of thermal noise on recovering one-point statistics with the MWA Phase I Core and HERA, we calculate thermal uncertainties for the statistics derived from our noiseless simulations. Our derivation of the thermal uncertainties follows \citet{Watkinson:2014jv}. The thermal uncertainty in interferometric measurements ($\Delta T^N$) is associated with the instrument and observing parameters \citep{2006PhR...433..181F}.
\begin{equation}\label{eq:noise1}
    \Delta T^N = \frac{T_{sys}}{\eta_f\sqrt{\Delta\nu t_{int}}},
\end{equation}

Here, $\eta_f = A_{tot} / D_{max}^2$ is the array filling factor, defined as the ratio of the total effective area ($A_{tot}$) and the maximum baseline length ($D_{max}$) of the array. $\Delta\nu$ is the spectral channel size, and $t_{int}$ is the integration time of the observation. $T_{sys}$ is the system temperature of the array, which is dominated by the Galactic synchrotron radiation in redshifted 21~cm observations below 200~MHz. This becomes the main contribution to the noise, giving $T_{sys} \approx T_{sky}=180 (\nu/180\mathrm{MHz})^{-2.6}\mathrm{K}$.  Using this assumption, Equation~\ref{eq:noise1} can be rewritten to estimate the thermal uncertainty in 21~cm observations,
\begin{equation}\label{eq:noise2}
\begin{split}
    \sigma_{\mathrm{noise}} &= 2.9\,\mathrm{mK} 
    \left(\frac{10^5\,\mathrm{m}^2}{A_{tot}}\right)
    \left(\frac{10\,\mathrm{arcmin}}{\Delta\theta}\right)^2 \\
    &\times \left(\frac{1+z}{10.0}\right)^{4.6}
    \sqrt{\left( \frac{1\,\mathrm{MHz}}{\Delta\nu} \frac{100\mathrm{h}}{t_{int}}\right)}
\end{split}
\end{equation}

Assuming that independent Gaussian random noise with zero mean and standard deviation $\sigma_{\mathrm{noise}}$ is added to every independent ``beam'' pixel in the simulated maps, the estimator variance of each p-th order statistical moment resulting from the noise can be calculated and later propagated to variance, skewness and kurtosis. The uncertainty for each statistic is then just the square root of the estimator variance. The noise propagation is described in detail in the Appendix, where we re-derive the equations given by \citet{Watkinson:2014jv}, along with an additional derivation for the kurtosis. 

In summary, the estimator variance of the 2nd, 3rd and 4th moments can be described by,
\begin{gather}
    \label{eq:m2_var}
    V_{\hat{m}_2} = V_{S_2} = \frac{2}{N_{beam}}
        (2 m_2 \sigma_{noise}^2 + \sigma_{noise}^4), \\
    \label{eq:m3_var}
    V_{\hat{m}_3} = \frac{3}{N_{beam}}
        (3 m_4 \sigma_{noise}^2 + 12 m_2 \sigma_{noise}^4 
        + 5 \sigma_{noise}^6), \\
    \label{eq:m4_var}
    \begin{split}
    V_{\hat{m}_4} = \frac{8}{N_{beam}}
    (2 m_6 \sigma_{noise}^2 + 21 m_4 \sigma_{noise}^4 \\
    + 48 m_2 \sigma_{noise}^6 + 12 \sigma_{noise}^8),
    \end{split}
\end{gather}
and the estimator variance of skewness and kurtosis are as follows,
\begin{gather}
    \label{eq:skew_var}
    V_{S_3} \approx \frac{1}{(m_2)^3}V_{\hat{m}_3}
        + \frac{9}{4}\frac{(m_3)^2}{(m_2)^5}V_{\hat{m}_2}
        - 3\frac{m_3}{(m_2)^4}C_{\hat{m}_2\hat{m}_3}, \\
    \label{eq:skew_var}
    V_{S_4} \approx \frac{1}{(m_2)^4} V_{\hat{m}_4}
        + 4 \frac{(m_4)^2}{(m_2)^6} V_{\hat{m}_4} 
        - 4 \frac{m_4}{(m_2)^5} C_{\hat{m}_2\hat{m}_4}. 
\end{gather}
Here, $N_{beam}$ is the number of independent PSF samples in the data, which is limited by the angular resolution of the telescope. Because our maps are oversampled, we calculate the number of independent PSF per pixel from the ratio of the pixel and the beam area, $f_{s}=\Delta\Omega_{pix} / \Delta\Omega_{beam}$, and multiply this factor to the number of pixels in our sample to obtain $N_{beam}$. $C_{\hat{m}_2\hat{m}_3}$ and $C_{\hat{m}_2\hat{m}_4}$ are the estimator covariance of the 2nd and 3rd moments, and the 3rd and 4th moments respectively. The estimator 
covariance can be derived in the same manner as the estimator variance of the moments resulting in,
\begin{gather}
    C_{\hat{m}_2\hat{m}_3} = \frac{6}{N_{beam}}m_3\sigma_{noise}^2, \\
    C_{\hat{m}_2\hat{m}_4} = \frac{4}{N_{beam}}
        (2 m_4 \sigma_{noise}^2 + 9 m_2 \sigma_{noise}^4
        + 3 \sigma_{noise}^6). 
\end{gather}

\begin{figure}[t]
    \centering
    \includegraphics[width=0.5\textwidth]{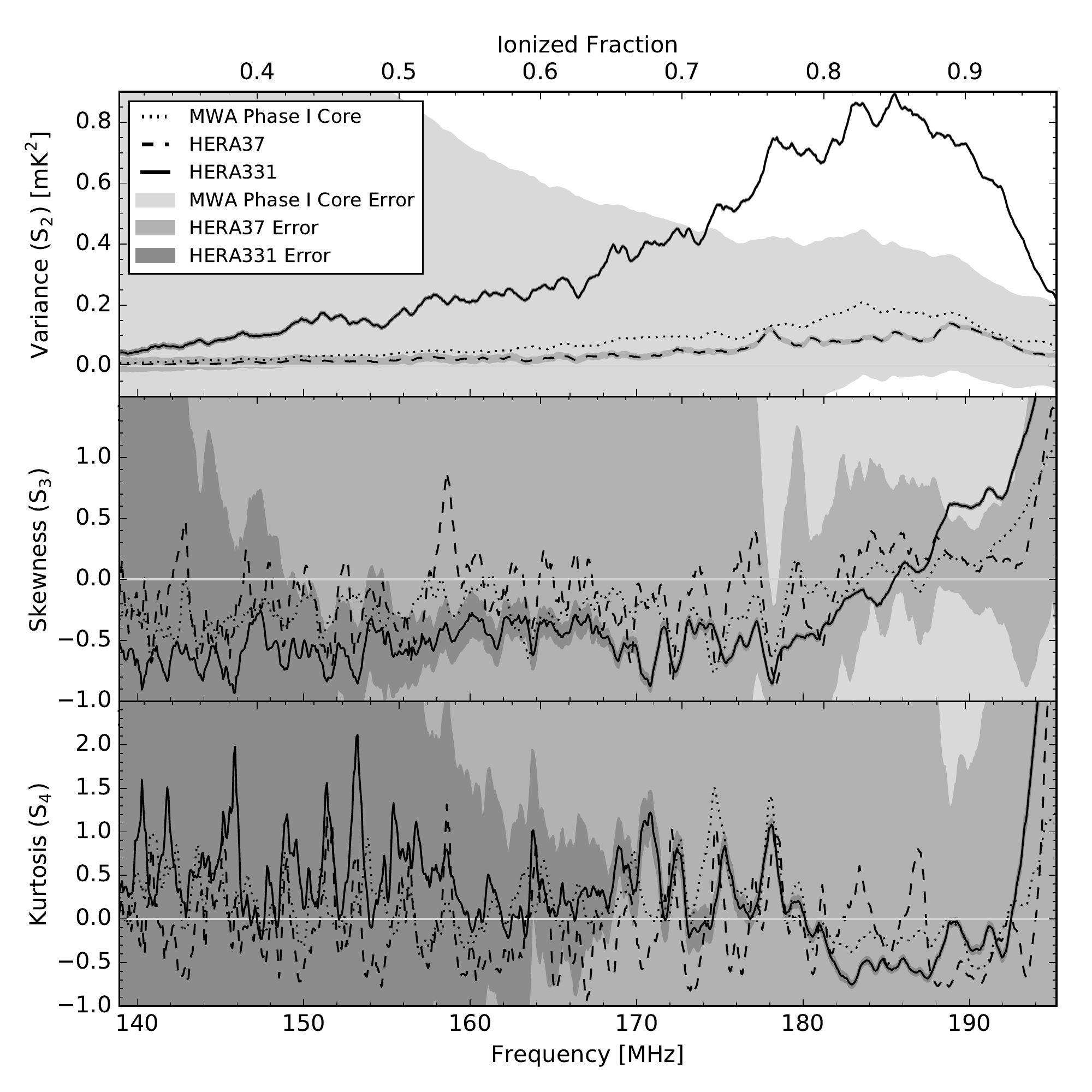}
    \caption{Variance (top), skewness (middle) and kurtosis (bottom) with estimated uncertainty from thermal noise. The dotted, dashed and solid lines show statistics derived from MWA Phase I Core, HERA37 and HERA331 simulations using Gaussian kernels as PSFs. Shaded regions are estimated thermal uncertainty for each telescope, starting from lighter to darker from MWA Phase I Core to HERA. Note that the uncertainty for HERA331 is not visible in the variance plot and only becomes prominent in skewness and kurtosis below ~170 MHz. All simulations here are performed with 80kHz frequency channels. Frequency windowing/binning will significantly improve telescope sensitivity as discussed in the main text and shown in Figures~\ref{fig:var}, \ref{fig:skew} and \ref{fig:kurt}.}
    \label{fig:thermal}
\end{figure}

Figure~\ref{fig:thermal} shows statistics derived from the Gaussian simulation for all three telescopes along with corresponding estimated thermal uncertainty.  We begin with this figure showing the thermal uncertainty for the 80 kHz (roughly native) spectral resolution of the arrays.   In this case, the narrow spectral channels result in substantial thermal noise.  We address frequency binning to reduce the uncertainty in the next section. The higher angular resolution of HERA331, compared to HERA37 and the MWA Phase I core, smooths out less signal, especially early in reionization when feature sizes are smaller, leaving more outliers in the field. This results in an overall higher observed variance that peaks at lower frequency, specifically at 185.475 MHz compared to at 188.835 MHz for HERA37. Similarly, skewness and kurtosis derived from the HERA331 simulation show more fluctuation, especially earlier in reionization, due to more outliers being left in the observed field. The overall trend of skewness and kurtosis, however, remain the same. In contrast, the larger PSFs of both HERA37 and the MWA Phase I Core smooth out more signal, resulting in overall lower variance and fewer spikes in skewness and kurtosis..

The thermal uncertainty rapidly decreases as the reionization progresses regardless of the statistics derived from the telescopes. This is because the synchrotron radiation, which dominates the system temperature, is brighter at lower frequencies. In terms of telescope sensitivity, it is clear from Figure~\ref{fig:thermal} that the MWA Phase I Core will not have enough sensitivity to detect any statistical features with the current observational parameters using 80 kHz spectral resolution, where as HERA331 array will be able to detect most features in all statistics with high signal-to-noise ratio (SNR) due to the much improved sensitivity of the array. Even at 80 kHz spectral resolution, we see that HERA37 will also be able to detect variance and place some decent limits on skewness and kurtosis near the end of reionization despite a comparable angular resolution to the MWA Phase I Core due to significantly larger collecting area. We show in Section~\ref{sec:bandwidth} that frequency binning will significantly improve telescope sensitivity for all statistics, including enable the MWA Phase I Core to detect features in the variance with sufficient SNR.

\subsection{Frequency Binning and Windowing}
\label{sec:bandwidth}

\begin{figure}
    \centering
    \includegraphics[width=0.5\textwidth]{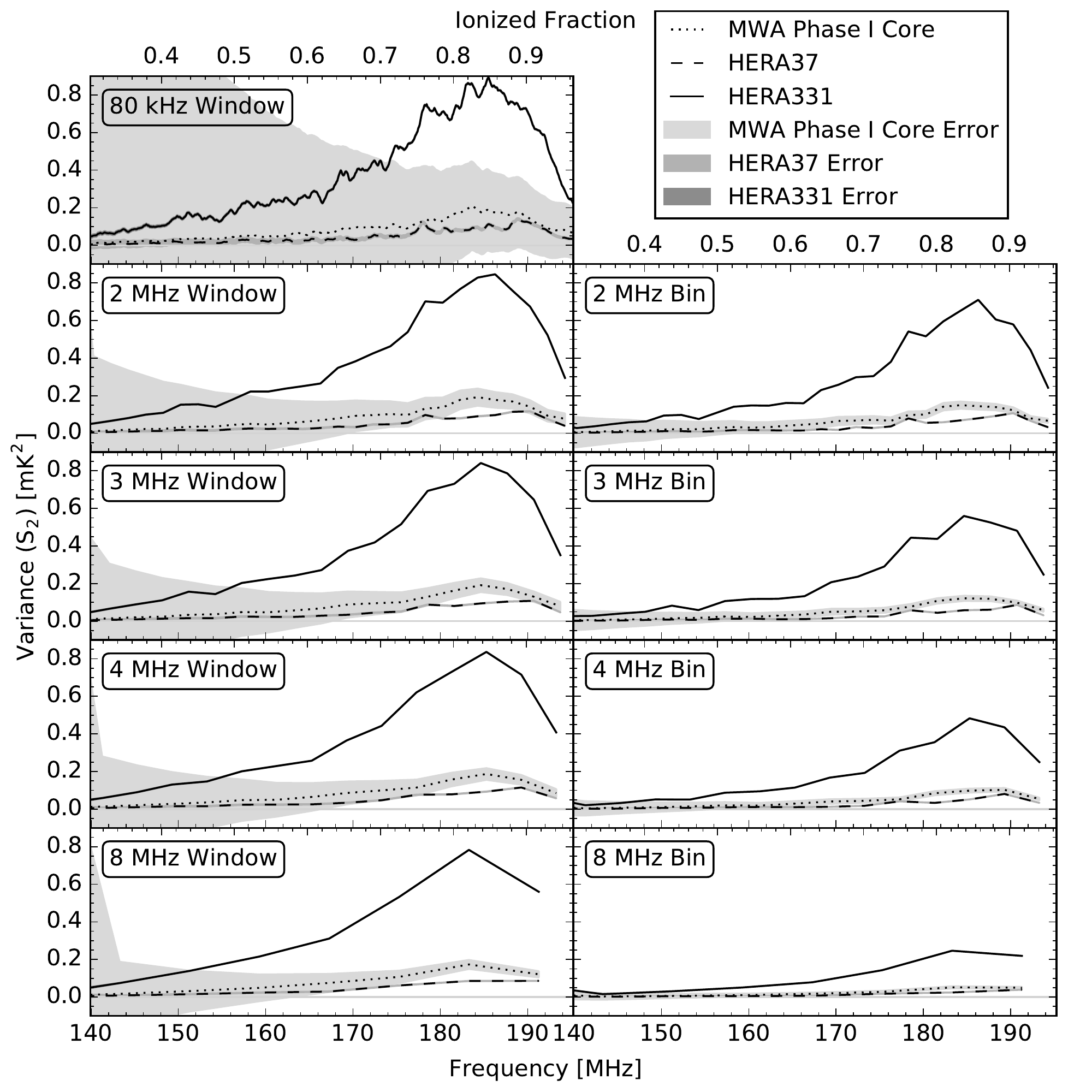}
    \caption{Variance evolution with uncertainty derived from the MWA Phase I Core (dotted), HERA37 (dashed) and HERA331 (solid) simulations with $\Delta \nu=$~2, 3, 4 and 8~MHz for frequency binning (right column) and frequency windowing (left columns) cases. The top left panel show results from the raw spectral channel of 80~kHz window. Note that uncertainty for HERA331 is not visible.}
    \label{fig:var}
\end{figure}
\begin{figure}
    \centering
    \includegraphics[width=0.5\textwidth]{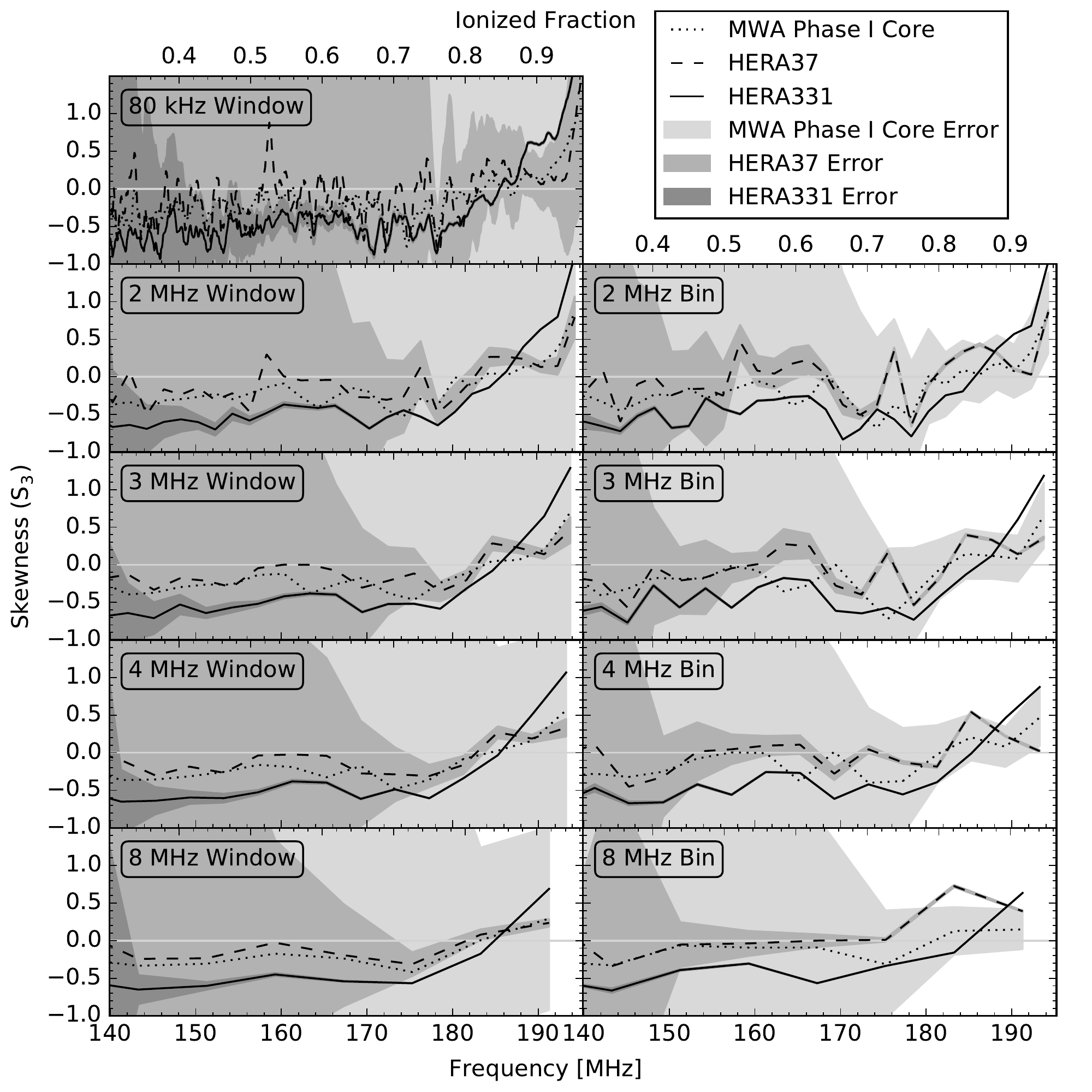}
    \caption{Same as Figure~\ref{fig:var}, but for skewness.}
    \label{fig:skew}
\end{figure}
\begin{figure}
    \centering
    \includegraphics[width=0.5\textwidth]{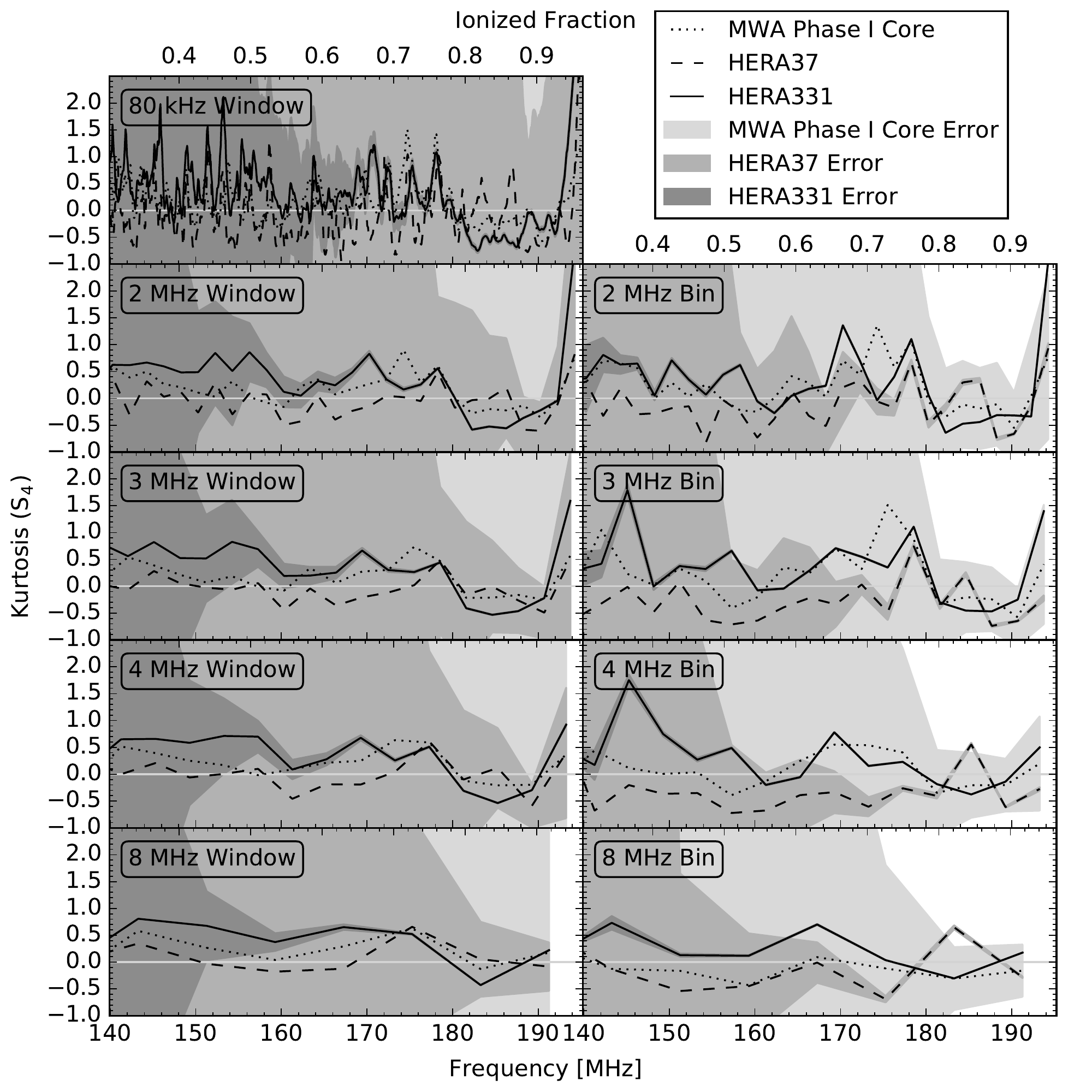}
    \caption{Same as Figure~\ref{fig:var}, but for kurtosis.}
    \label{fig:kurt}
\end{figure}

It is evident from Equations~\ref{eq:noise2}--\ref{eq:m4_var} that the thermal uncertainty on the one-point statistics is sensitive to both the spectral channel size ($\Delta \nu$) of the observed maps and the number of samples ($1/N_{beam}$) used to calculate a given estimate of a statistic.  Both of these parameters can be tuned to achieve the best-possible signal-to-noise ratio (SNR) for a particular instrument and 21~cm signal model combination.  We explore the benefits of varying spectral channel size through frequency binning (averaging) of raw maps, as well as the effects of increasing the number of samples per estimate by grouping neighboring maps into the calculation of a one-point statistic without frequency averaging.  We term the latter process frequency windowing.

From Equation~\ref{eq:noise2}, we see that increasing the spectral channel size, $\Delta \nu$, reduces the thermal noise in a given observation.  Both MWA and HERA have relatively narrow raw spectral channels ($\Delta \nu_{raw}<100$~kHz) yielding high thermal noise in raw maps.  However, we can bin observed maps in neighboring frequency channels to yield larger effective $\Delta \nu$ and lower thermal noise per binned map.  Frequency binning does have potentially detrimental effects.  It comes at the expense of reducing the number of samples that can be used to chart the evolution of the signal, and it can also reduce the observed 21~cm signal amplitude if it is performed over a range larger than the coherence length of the signal. Unexpected amplitude gain or loss can also occur as we shall see in  the results.

Frequency windowing uses maps created at the raw spectral channel width (80~kHz in our cases), but calculates statistics using multiple maps for each estimate.  The motivation for this method is the inclusion of more samples into each statistical calculation.  The noise per pixel is unchanged from the raw maps as is the underlying signal contribution to each pixel, but using more pixels per estimate lowers the overall uncertainty on the estimate as shown in Equation~\ref{eq:m2_var}--\ref{eq:m4_var}.  The disadvantage is that frequency windowing has the potential to dilute evolution in the signal similar to frequency binning. 
  
To investigate the trade-off between the improvement in thermal uncertainty and the potential loss of signal of the two methods, we group our MWA and HERA Gaussian simulations into trial cases of different total spectral window size, spanning 1-30~MHz with 1~MHz increment between each case. Then, we measure the variance, skewness and kurtosis, with and without binning the simulation sets, using the same method as described in Section~\ref{sec:1point_stats_definition}, and calculate the corresponding thermal uncertainty as above.

Figures~\ref{fig:var},~\ref{fig:skew}~and~\ref{fig:kurt} show the variance, skewness and kurtosis and their uncertainty derived from the MWA (dotted), HERA37 (dashed) and HERA331 (solid) simulations with $\Delta \nu=$~2, 3, 4 and 8~MHz. The right and left columns show results from the frequency binning and frequency windowing trials respectively. In addition, the top left panel in all three figures show results from our raw spectral channel size (0.08~MHz window), same as those in Figure~\ref{fig:thermal}. We select these particular frequencies to show several trends that are apparent in our results.

\begin{figure}[t]
    \includegraphics[width=0.5\textwidth]{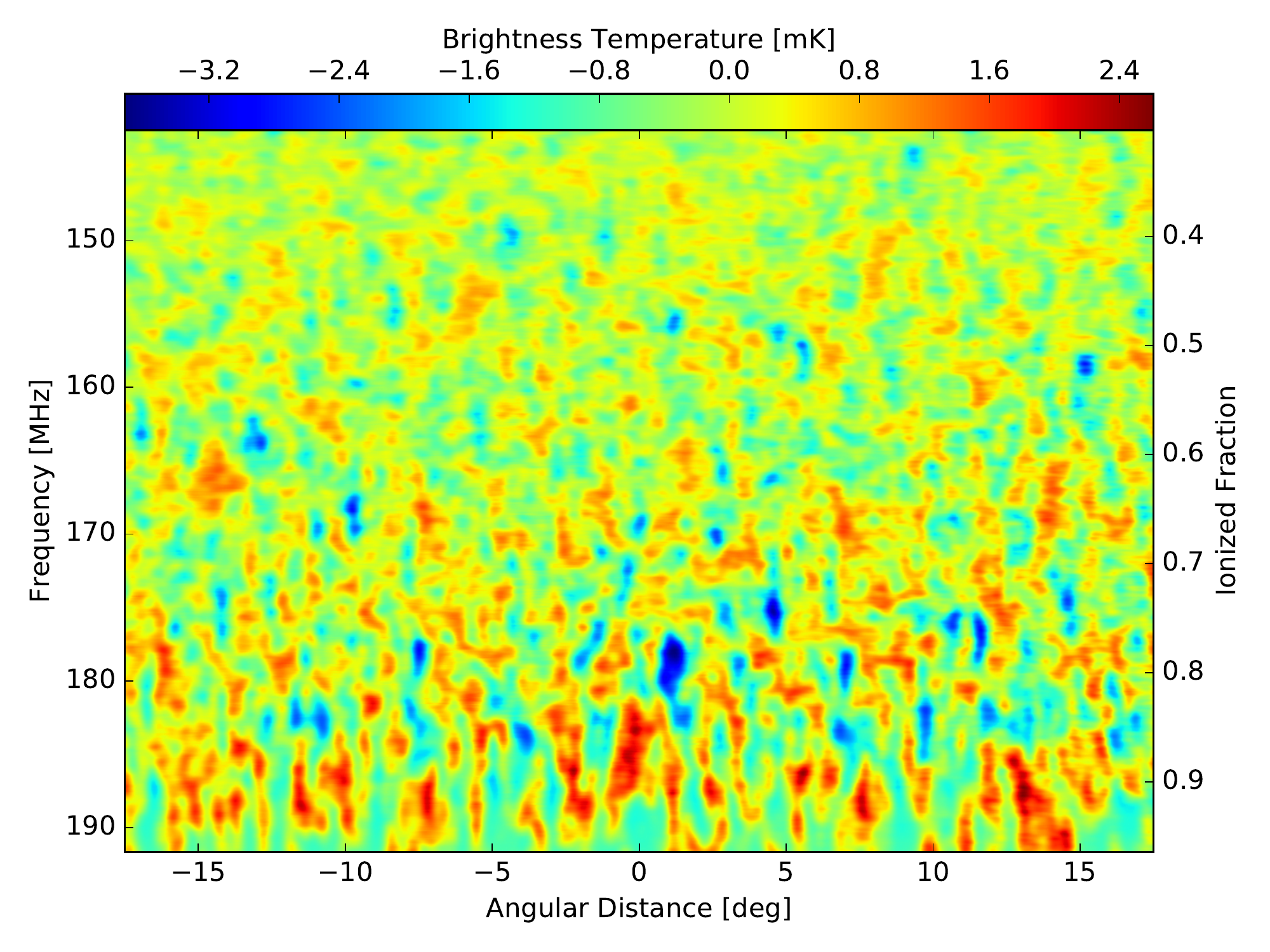}
    \caption{Lightcone map from HERA331 simulation, illustrating the evolution of the 21~cm signal as a function of ionized fraction and frequency.}
    \label{fig:lightcone}
\end{figure}

Frequency binning produces similar effects to the PSF smoothing but along the frequency axis. Binning reduces observed statistical variance particularly at the larger bin sizes by smoothing out the signal over the frequency scale corresponding to the bin sizes. To illustrate this effect, a lightcone map showing the evolution of signal in the HERA331 simulated observations is shown in Figure~\ref{fig:lightcone}. The typical sizes of the signal along the frequency axis grow as reionization progresses, reaching the maximum size of $\sim5$~MHz near the end of reionization. Binning the signal with bin size greater than this size scale will significantly reduce the amplitude of signal since there are few features that are larger than this size. The observed variance declines significantly between 4~and 8~MHz bins. 

Skewness and kurtosis do not suffer overall amplitude reduction from the binning due to both statistics being sensitive to the relative shape of distribution, not the amplitude of the signal. However, binning smooths out the extremely low and high regions in the maps to neighboring frequency channels within the bin. Thus, the binned skewness and kurtosis appear to have fewer occasional fluctuation throughout. Binning can also reduce or amplify the overall amplitude of these regions depending on their appearances in the neighboring frequency bins and the bin location. A large peak appears in the kurtosis near 145 MHz when the bin size is 3--4~MHz but disappears when the bin size is smaller or larger. Interestingly, this peak appears only in the HERA331 simulation where the angular resolution is ~0.5 degree but not the HERA37 or MWA Phase I Core simulations where the angular resolution is ~1.5 degree. This implies a correlation between angular resolution and the amplitude of the kurtosis peaks. We explore its origin in the next section.

Compared to frequency binning, frequency windowing simply adds more samples to the statistics. Windowing over a large scale does not directly reduce the signal amplitude but rather convolves the intrinsic variance across a range of redshifts. Thus, we see that the variance amplitudes remain at similar levels regardless of window size. Apart from retaining the variance amplitude, frequency windowing retains large numbers of pixels in each estimate, reducing the susceptibility to sample variance within each map, and allowing the statistics to evolve smoothly with redshift. A good example is the skewness in Figure~\ref{fig:skew} in which frequency binning results in larger variation in the skewness from one redshift to the next while frequency windowing produces skewness that smoothly evolves.

\begin{figure}[t]
    \includegraphics[width=0.5\textwidth]
        {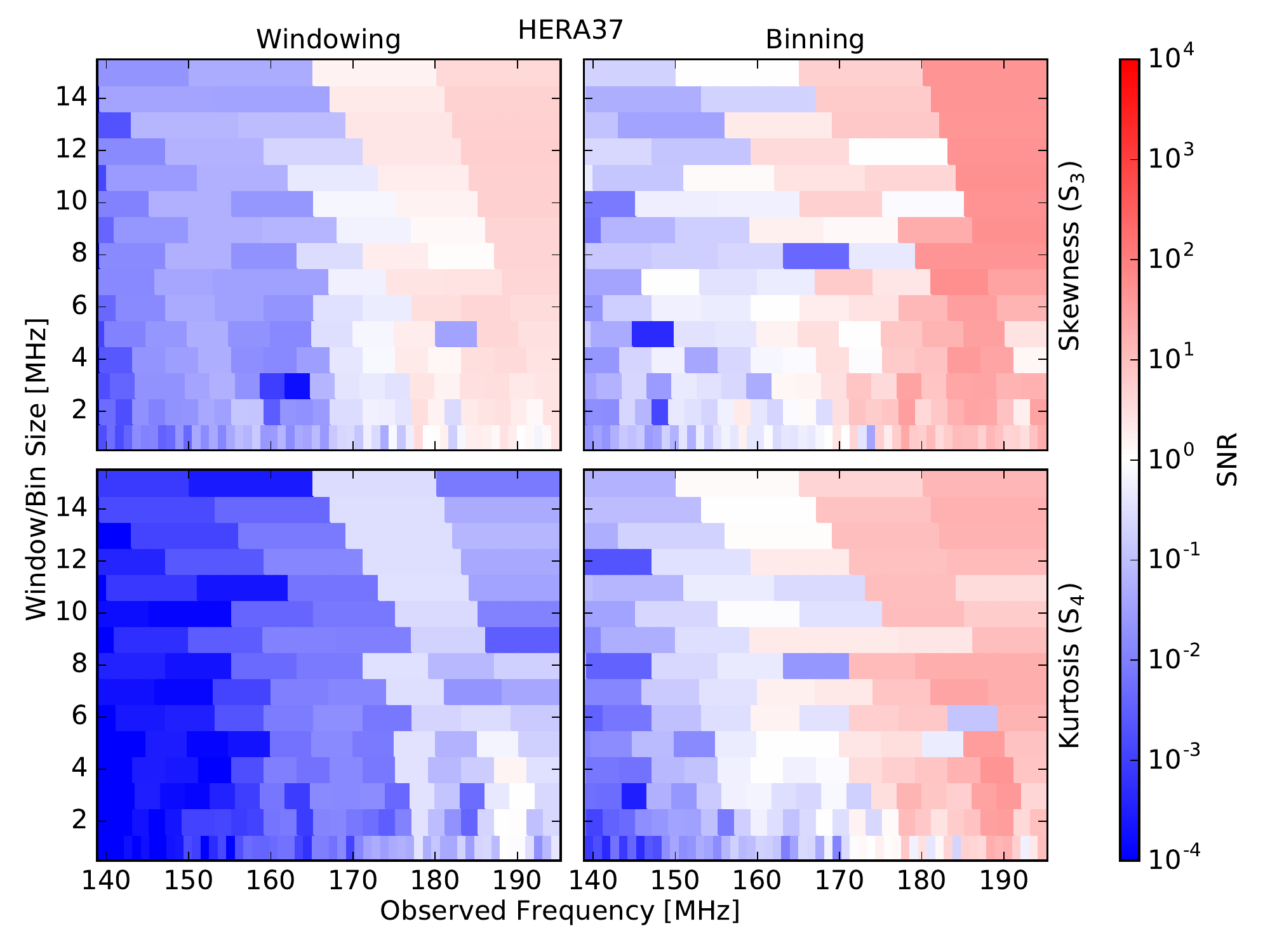}
    \caption{SNR of different frequency windowing (left column) and frequency binning (right column) cases for HERA37 with the top panels showing SNR for skewness and bottom panels for kurtosis. The colorbar is normalized to white, blue and red colors at SNR=1, $<1$ and $>1$ respectively. Each block in the diagram depicts the bin/window at respective bin/window sizes}
    \label{fig:snr_hera37}
\end{figure}

\begin{figure}
    \includegraphics[width=0.5\textwidth]
        {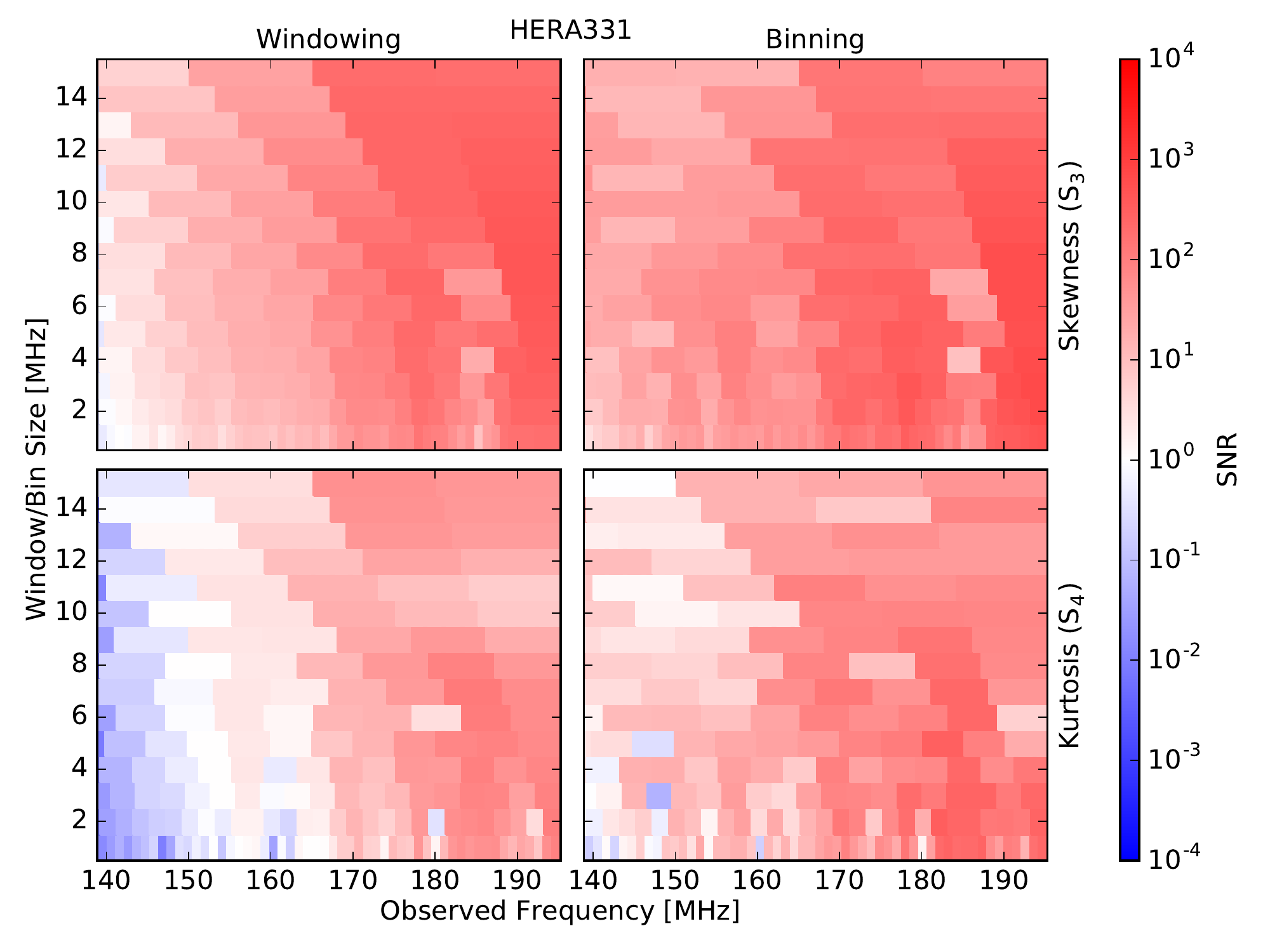}
    \caption{Same as Figure~\ref{fig:snr_hera37}, but for HERA331.}
    \label{fig:snr_hera331}
\end{figure}

An important question to ask is whether we should use frequency binning or frequency windowing and at what bin or window size for each array. For the MWA Phase I Core, Figure~\ref{fig:var} suggests that frequency binning is preferable since it tends to yield greater signal to noise. For HERA, an overall look at the signal-to-noise ratio for all cases provides a more robust answer. Figures~\ref{fig:snr_hera37} and \ref{fig:snr_hera331} show color charts that depict the SNR of skewness and kurtosis, derived from the HERA37 and HERA331 simulations, as a function of bin/window size. Each block in this diagram represent the bin/window for each cases, with windowing cases in the left column, and binning cases in the right column. The scale is normalized such that the color is white at SNR=1, and becomes bluer and redder at SNR$<1$ and SNR$>1$ respectively. Based on the chart, we find that statistics derived from the frequency binning approach have lower uncertainty than the frequency-windowed counterparts for the same bin/window size. This should be true for all other arrays, including the MWA Phase I Core. A mathematical explanation is that the variance for p-th order moment is approximately proportional to the noise uncertainty to the power of p$^{2}$ and the reciprocal of the number of beam, $V_{\hat{m}_p} \propto \sigma_{noise}^{2p} / N_{beam}$. Equation~\ref{eq:noise2} implies that $\sigma_{noise} \propto 1 / \sqrt{\Delta \nu}$, leading to $V_{\hat{m}_p} \propto 1 / [(\Delta \nu)^p N_{beam}]$. Binning $X$ raw channels will reduce the moment uncertainty by $X^p$ while windowing $X$ raw channel will reduce the moment uncertainty by just $X$. Thus, we suggest that spectral binning be use to detect features in the statistics. Figures~\ref{fig:snr_hera37} and \ref{fig:snr_hera331} also suggest that a bin size of $\sim4-6$~MHz be used as averaging over wider frequency range does not significantly improve the signal-to-noise. This is in agreement with the previously discussed point that features reach the the maximum size of $~5$~MHz along the frequency axis, and averaging over a range larger than this number would only reduce the signal.


\subsection{Correlation in the Observed Kurtosis Peaks}\label{sec:kurtosis}

Throughout this work, we see occasional spikes that rise above the fluctuations caused by sample variance in the kurtosis. We find that these spikes occur when the observed field contains a few large regions of extremely low or high brightness temperature (relative to the mean at that frequency) (see Section~\ref{sec:1point_stats_features}). These outlier regions add more density to the tails of the PDFs, resulting in higher kurtosis. Binning multiple maps of neighboring frequencies that contain these outlier regions will smooth them over the frequency bin, resulting in a broader peak in the kurtosis. The amplitude of the peak will be reduced or amplified, depending on whether the size of the outlier regions along the frequency axis matches the frequency bin size. Thus, we see kurtosis peaks at 145.115 MHz and 178.555 MHz in the 2 and 3 MHz bin cases, but not in other cases. These particular peaks, however, only appear in the kurtosis derived from HERA331 simulation, but not from other telescope simulations (see Section~\ref{sec:bandwidth}. We also see more peaks in the kurtosis derived from the un-binned maps of HERA331 simulation in comparison to un-binned maps from other telescopes (see ~\ref{sec:hera}). These imply a correlation between kurtosis peaks, the angular resolution of PSF and the average bin size.

To understand more about this correlation, we perform additional simulations for intermediate HERA array sizes, specifically with 61, 91, 127, 169, 217 and 271 antennas, corresponding to approximately 1.1, 0.9, 0.8, 0.7, 0.6, 0.5 degree angular resolutions at 150 MHz, to give us more coverage over a range of angular resolution. We then recalculate the kurtosis from maps of all additional configurations, binning and windowing the maps with 1-30 MHz bin size as before. Noise parameters are adjusted to match the PSF resolutions.

\begin{figure*}[t]
    \includegraphics[width=\textwidth]{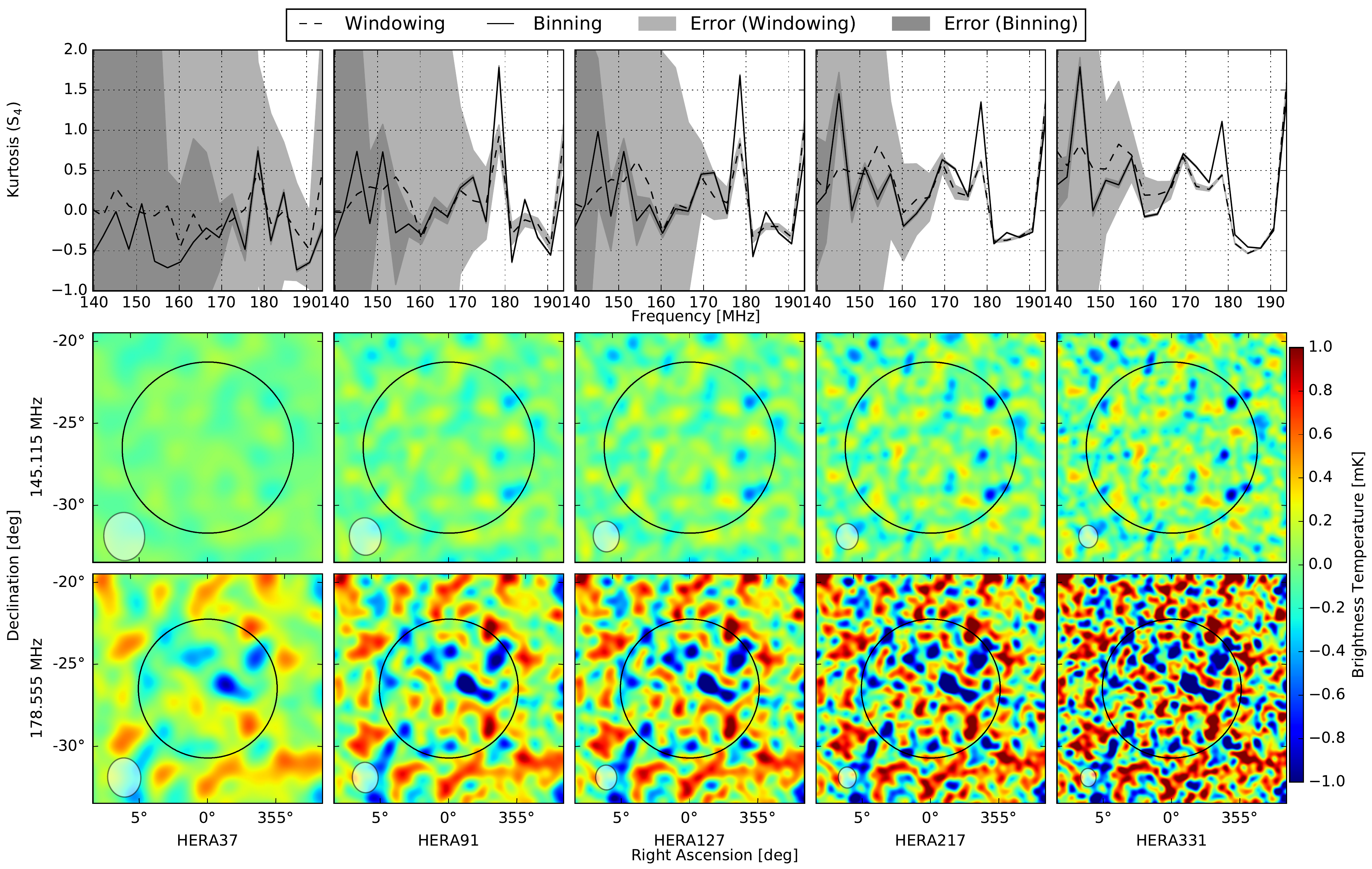}
    \caption{Illustrating the correlation between amplitudes of the kurtosis peaks, the telescope PSF resolution, and extremely high and low brightness temperature regions. The top row shows kurtosis derived from the 3~MHz windowed maps (dotted lines) and binned maps (solid lines) with lighter and darker gray shade representing the thermal uncertainty of the two cases. The bottom two rows are maps at 145.115 MHz (2nd row) and 178.55 MHz (3rd row) from simulation of HERA arrays with 37, 91, 127, 217 and 331 antennas (left to right), showing growth of the extreme regions in the fluctuations. A white shaded circle at the bottom left corner of each panel represents the size of the PSF for each array, whereas a black circle shows the field of view. Only pixels inside the field of view are used in the calculation.}
    \label{fig:kurt_maps}
\end{figure*}

As expected, we see the correlation between the kurtosis peaks and the angular resolution of the telescopes. Figure~\ref{fig:kurt_maps} illustrates the relation. The top panel shows the kurtosis derived from the 3~MHz windowed and binned maps of HERA37, HERA91, HERA127, HERA217 and HERA331 simulations (left to right). Other array configurations are left off for simplicity. The bottom two panels show maps from the simulation at 145.115 MHz (2nd row) and 178.555 MHz (3rd row) from the respective telescopes. In addition, the PSF size of each telescope is drawn as a white ellipse at the bottom left corner of each map. It is clear from the diagram that the amplitude of peak at 145.115 MHz rises as the sizes of outlier regions in the map start to match the size of the PSF. Since the size scale of the signal at 145.555 MHz is small, the kurtosis peak reaches its maximum when observed with the matched PSF of HERA331. Similarly, the peak at 178.555 MHz reaches its maximum when observed with HERA91 where the larger PSF of the telescope matches the typical size scale at that frequency. Although not shown here, we see the same trend in the kurtosis derived from maps of other bin sizes. For example, the two peaks near 171 and 179 MHz in the kurtosis derived from 2 MHz binned maps (2nd row, right, in Figure~\ref{fig:kurt}) reach their maximum when observed with HERA217 and HERA91 respectively, rising and falling according to the sizes of the PSF. The peak near 145 MHz in the 4 MHz binned maps also rises as the PSF resolution increases, reaching its maximum when observed with HERA331 similar to the peak at 15.115 MHz in the 3 MHz bin case.

\section{Conclusion}\label{sec:conclusion}

For this work, we have developed realistic simulations of EoR observations that produce dirty maps of 21~cm brightness temperature intensity fluctuations. Our simulation incorporates full-sky 21~cm models and realistic beam response. We build full-sky 21~cm models with a tile-and-grid method that projects and interpolates small 21~cm simulation cubes to lightcone gridded HEALPix maps. We base our simulations on the MWA Phase I Core configuration. Simple Gaussian smoothing is also performed for a comparison with additional PSF and noise parameters from HERA37 and HERA331 array. We measure the PDF, variance, skewness and kurtosis from the simulated maps and study the effect of frequency binning and windowing on the statistics. We also study the correlation between the size of the array point spread function and the peaks in observed kurtosis by performing additional simulations based on different HERA configurations. Uncertainty from thermal noise is mathematically derived and propagated to the statistics. Errors from foreground contamination and other systematics are ignored in this work, postponing them to future works.

Base on our findings we conclude with the following.

\begin{itemize}
    \item Gaussian PSF kernels can reasonably approximate the full instrument simulations for arrays with densely-pack antenna configuration such as the MWA Core. High correlation is maintained between statistics derived from simulations with Gaussian and full PSFs although sidelobe confusion in the full PSF can scale the measured statistics by as much as 40-60$\%$. A transfer function could potentially be used to rescale the signal, but more studies with various array types and 21~cm models are necessary. 
    \item Apart from spatially smoothing the signal by the PSF shape, the instrument acts analogously to a match filter that emphasizes a spatial size scale that match the angular resolution of the PSF. The observed variance will have lower overall amplitude and reach the maximum peak at a different redshift depending on the angular resolution of the PSF. Skewness and kurtosis remains near zero for most of reionization due to the underlying distribution becoming more Gaussian-like from PSF smoothing, except near the end of reionization where the intrinsic signal is highly bi-modal.
    \item Measuring the statistics over a limited field of view introduces sampling variance, resulting in statistics that oscillate over a range of redshifts as opposed to the smoothly varying curves derived from the models. Coupling with the size-scale emphasis of the PSF, this can result in strong peaks that rise above variation from sampling variance as we have seen in the kurtosis measurements throughout this work. 
    \item Similarly, frequency binning smoothes and emphasizes certain size scales along the frequency axis apart from improving the signal-to-noise of the measurement. Thus, observed variance will have lower overall amplitude if measured from frequency binned maps, and certain peaks can appear in the kurtosis derived from a specific bin size but not another. In contrast, frequency windowing (grouping neighboring frequency channel without averaging) preserves the variance of the signal and reduces sampling variance in the observed statistics due to higher statistical samples in the windowed data. Frequency windowing, however, is less effective in reducing thermal noise.
    \item The origin of the peaks in the observed kurtosis can be physically explained by considering the distribution of regions with extremely low or high brightness temperature, relative to the mean temperature at the redshift. When these regions dominate the observing field, observed kurtosis will increase due to higher density being added to the tails of the PDF of the measured signal. Similarly, blurring of these regions due to mismatched bin size or PSF size, will reduce the amplitude of the peaks. Hence, we see peaks only at specific frequency binning and angular resolution combinations. The correlation between the kurtosis peaks, frequency bin size and angular resolution of the telescopes could be very beneficial in quantifying 21~cm signal. A similar study to this work performed over wider range of these variables can potentially result in an empirical relationship that quantifies the statistical correlation between 21~cm model and observed number of large kurtosis and skewness features.  Although these regions may not be the exact representations of the ionized or overdense regions, due to the instrumental smoothing of the complex underlying ionization structure, they can be statistically related as demonstrated in \citet{2015ApJ...800..128B}.
    \item In terms of detectability, our study suggests that the MWA Phase I Core should be able to detect the variance peak with SNR$\sim2$, while HERA, even with the smaller 37 array, will be able to detect all studied statistics with high sensitivity throughout the reionization redshifts. Frequency binning significantly improves the sensitivity for all detections, and our study suggests that the bin size of $\sim4-6$~MHz be used to optimize SNR.
\end{itemize}

\acknowledgements
We thank Matt Mechtley, Ian Sullivan, Leonid Benkevitch and Randall Wayth for their valuable inputs on the software development for this work. This work was supported by the U.S. National Science Foundation (NSF) through awards AST-1109156. PK acknowledges support from the Royal Thai Government Scholarship provided by the Development for Promotion of Science and Technology Talent Project of Thailand. DCJ acknowledges support from an NSF Astronomy and Astrophysics Postdoctoral Fellowship under award AST-1401708. NT acknowledges support from NSF awards AST-1440343 and AST-1410719.

\appendix
\label{appendix}

\section{Uncertainty Propagation}

The method of uncertainty propagation for one-point statistics is first described in \citet{Watkinson:2014jv}. Here, we summarize and expand on their work, deriving uncertainty propagation for the kurtosis in addition to variance and skewness.

To recap, for a map with pixel value $x_i$, mean $\overline{x}$ and $N_{pix}$ pixels, the p-th central moment of the map is defined as, 
\begin{equation}
    m_p = \frac{1}{N_{pix}}\sum_{i=0}^{N_{pix}} (x_i - \overline{x})^p.
\end{equation}
Then, the variance, skewness and kurtosis are standardizations of 2nd, 3rd and 4th central moments as follow.
\begin{align}
    \text{variance:\quad} &S_2 = m_2, \\
    \text{skewness:\quad} &S_3 = \frac{m_3}{(m_2)^{3/2}}, \\
    \text{kurtosis:\quad} &S_4 = \frac{m_4}{(m_2)^2} - 3, 
\end{align}

Assuming that every pixel $x_i$ in a 21~cm brightness temperature fluctuation map represents an independent brightness temperature fluctuation ``pixel'' $\delta T_i$, we can simply substitute $x_i=\delta T_i$ and $\overline{x}=\overline{\delta T}$ to compute the ``true'' moments of the brightness temperature fluctuation. 

Adding noise $n_i$ with standard deviation $\sigma_i$ to the true signal, each pixel now consists of the true signal plus the noise, $x_i = \delta T_i + n_i$, and the noise will bias the moments. An unbiased estimator for the p-th moments $\hat{m}_p$ can be estimated by averaging the moment equations over noise realization. Assuming that the noise is Gaussian and independent on different pixels, the averaged noise terms can be rewritten as  functions of standard deviation of Gaussian noise using the Gaussian moment identity,
\begin{equation}\label{eq:gauss_id}
    \langle n_i^l \rangle = 
    \begin{cases}
    (1)(3)(5)\dotsm(l-1)\sigma_i^l & \text{if } l \text{ is even}\\
    0  & \text{if } l \text{ is odd}\\
    \end{cases}
\end{equation}
where the angle bracket designates an average. In addition to Equation~\ref{eq:gauss_id} and the identities given in the Table A1 in \citet{Watkinson:2014jv}, Table~\ref{tab:noise} in this work provide more identities necessary for the derivation of the kurtosis uncertainty.

The uncertainty for each statistic that will arise from the added noise can be determined by calculating the estimator variance and covariance of the unbiased estimator of the moments,
\begin{gather}
    V_{\hat{m_p}} = \langle\hat{m_p}\hat{m_p}^{\dagger}\rangle 
        - \langle\hat{m_p}\rangle^2, \\
    C_{\hat{m}_p\hat{m}_q} = \langle\hat{m}_p\hat{m}_q\rangle 
        - \langle\hat{m}_p\rangle\langle\hat{m}_q\rangle.
\end{gather}
Since skewness and kurtosis are functions of the 2nd and 3rd, and 2nd and 4th moments, their estimator variance can be calculated by propagating the estimator variance of the moments to respective statistics using Taylor expansion,
\begin{equation}\label{eq:taylor}
\begin{split}
    V_{f(X,Y)} \approx 
    \left(\frac{\partial f}{\partial X}\right)^2 
    V_{X}
    + \left(\frac{\partial f}{\partial Y}\right)^2
    V_{Y} \\
    + 2 \left(\frac{\partial f}{\partial X}\right) 
    \left(\frac{\partial f}{\partial Y}\right)
    C_{X Y}
\end{split}
\end{equation}
Here, $f(X,Y)$ is a function of two non-independent variables $X$ and $Y$. In other word, $X=m_2$ and $Y=m_3$ for skewness, and $X=m_2$ and $Y=m_4$
for kurtosis.

Equations~\ref{eq:old1} to \ref{eq:old2} summarize results from \citet{Watkinson:2014jv}. Here, $\sigma_i$ is assumed to be equal to $\sigma_{noise}$ in Equation~\ref{eq:noise2} for all pixels. 
\begin{gather}
    \label{eq:old1}
    \hat{m}_2 = \frac{1}{N_{pix}} \sum_{i=0}^{N_{pix}}
        (\delta T_i - \overline{\delta T})^2 
        - \sigma_{noise}^2, \\
    \hat{m}_3 = \frac{1}{N_{pix}} \sum_{i=0}^{N_{pix}}
        (\delta T_i - \overline{\delta T})^3, \\
    C_{\hat{m}_2\hat{m}_3} = \frac{6}{N_{pix}}m_3\sigma_{noise}^2, \\
    V_{\hat{m}_2} = V_{Var} = \frac{2}{N_{pix}}
        (2 m_2 \sigma_{noise}^2 + \sigma_{noise}^4), \\
    V_{\hat{m}_3} = \frac{3}{N_{pix}}
        (3 m_4 \sigma_{noise}^2 + 12 m_2 \sigma_{noise}^4 
        + 5 \sigma_{noise}^6), \\
    V_{S_3} \approx \frac{1}{(m_2)^3}V_{\hat{m}_3}
        + \frac{9}{4}\frac{(m_3)^2}{(m_2)^5}V_{\hat{m}_2}
        - 3\frac{m_3}{(m_2)^4}C_{\hat{m}_2\hat{m}_3}. 
    \label{eq:old2}
\end{gather}

We follow the procedure in \citet{Watkinson:2014jv} and are able to confirm their results. In addition, we derive the estimator variance for kurtosis as follows. 

First the the 4th moment with added noise is constructed.
\begin{align}
    \hat{m}_4^{test} 
        &= \frac{1}{N_{pix}} \sum_{i=0}^{N_{pix}}
            (x_i - \overline{x})^4 \nonumber\\
        &= \frac{1}{N_{pix}} \sum_{i=0}^{N_{pix}}
            [(\delta T_i - \overline{\delta T}) 
            + n_i]^4.
\end{align}

Then, we average over the noise.
\begin{align}
    \langle \hat{m}_4^{test} \rangle
        &= \frac{1}{N_{pix}} \sum_{i=0}^{N_{pix}}
            [(\delta T_i - \overline{\delta T})^4 +
            4(\delta T_i - \overline{\delta T})^3 
            \langle n_i \rangle \nonumber\\
        &\quad+ 6(\delta T_i - \overline{\delta T})^2 
            \langle n_i^2 \rangle + 
            4(\delta T_i - \overline{\delta T}) 
            \langle n_i^3 \rangle + 
            \langle n_i^4 \rangle] \nonumber\\
        &= \frac{1}{N_{pix}} \sum_{i=0}^{N_{pix}}
            (\delta T_i - \overline{\delta T})^4 \nonumber\\
        &\quad+ 6 \frac{1}{N_{pix}} \sum_{i=0}^{N_{pix}}
            (\delta T_i - \overline{\delta T})^2 
            \sigma_{i}^2 + 3 \sigma_{i}^4 \nonumber\\
        &= \frac{1}{N_{pix}} \sum_{i=0}^{N_{pix}}
            (\delta T_i - \overline{\delta T})^4 \nonumber\\
        &\quad+ 6 m_2 \sigma_{noise}^2 + 3 \sigma_{noise}^4.
\end{align}

This implies that the unbiased estimator of the 4th moment is,
\begin{align}
    \hat{m}_4 
        &= \frac{1}{N_{pix}} \sum_{i=0}^{N_{pix}}
            (\delta T_i - \overline{\delta T})^4 
            - 6 m_2 \sigma_{noise}^2 
            - 3 \sigma_{noise}^4 \nonumber\\
        &= \frac{1}{N_{pix}} \sum_{i=0}^{N_{pix}}
            (\delta T_i - \overline{\delta T})^4 
            - \frac{3}{2}N_{pix}V_{\hat{m_2}}.
\end{align}
Next we derive the estimator variance of the 4th moment. We substitute $\mu_i=\delta T_i - \overline{\delta T}$ and $\kappa=3N_{pix}V_{\hat{m_2}}/2$ to simplify the derivation.
\begin{equation}
    \begin{split}
        V_{\hat{m}_4} = \biggl\langle \frac{1}{N_{pix}^2}
            \sum_{i=0}^{N_{pix}}\sum_{j=0}^{N_{pix}}
            \lbrack(\mu_i+n_i)^4-\kappa\rbrack \\
        \times \lbrack(\mu_j+n_j)^4-\kappa\rbrack
            \biggr\rangle - (m_4)^2.
    \end{split}
\end{equation}

Expanding this expression and moving the noise averaging brackets inside the summation gives,
\begin{align}
    V_{\hat{m}_4} = &\frac{1}{N_{pix}^2}
        \sum_{i=0}^{N_{pix}}\sum_{j=0}^{N_{pix}}
        \Big\lbrack\mu_i^4\mu_j^4 
        + 4\mu_i^4\mu_j^3\langle n_j \rangle \nonumber\\
        &+ 6\mu_i^4\mu_j^2\langle n_j^2 \rangle 
        + 4\mu_i^4\mu_j\langle n_j^3 \rangle 
        + \mu_i^4\langle n_j^4 \rangle 
        - \kappa\mu_i^4 \nonumber\\
        &+ 4\mu_i^3\mu_j^4\langle n_i \rangle
        + 16\mu_i^3\mu_j^3\langle n_i n_j \rangle 
        + 24\mu_i^3\mu_j^2\langle n_i n_j^2 \rangle\nonumber\\
        &+ 16\mu_i^3\mu_j\langle n_i n_j^3 \rangle 
        + 4\mu_i^3\langle n_i n_j^4 \rangle
        - 4\kappa\mu_i^3\langle n_i\rangle \nonumber\\
        &+ 6\mu_i^2\mu_j^4\langle n_i^2 \rangle 
        + 24\mu_i^2\mu_j^3\langle n_i^2n_j \rangle 
        + 36\mu_i^2\mu_j^2\langle n_i^2n_j^2 \rangle\nonumber\\ 
        &+ 24\mu_i^2\mu_j\langle n_i^2n_j^3 \rangle 
        + 6\mu_i^2\langle n_i^2n_j^4 \rangle
        - 6\kappa\mu_i^2\langle n_i^2\rangle \nonumber\\
        &+ 4\mu_i\mu_j^4\langle n_i^3 \rangle 
        + 16\mu_i\mu_j^3\langle n_i^3n_j \rangle 
        + 24\mu_i\mu_j^2\langle n_i^3n_j^2 \rangle\nonumber\\ 
        &+ 16\mu_i\mu_j\langle n_i^3n_j^3 \rangle 
        + 4\mu_i\langle n_i^3n_j^4 \rangle
        - 4\kappa\mu_i\langle n_i^3 \rangle \nonumber\\
        &+ \mu_j^4\langle n_i^4 \rangle 
        + 4\mu_j^3\langle n_i^4n_j \rangle 
        + 6\mu_j^2\langle n_i^4n_j^2 \rangle\nonumber\\ 
        &+ 4\mu_j\langle n_i^4n_j^3 \rangle 
        + \langle n_i^4n_j^4 \rangle
        - \kappa\langle n_i^4 \rangle \nonumber\\
        &- \kappa\mu_j^4
        - 4\kappa\mu_j^3\langle n_j \rangle
        - 6\kappa\mu_j^2\langle n_j^2 \rangle
        - 4\kappa\mu_j\langle n_j^3 \rangle \nonumber\\
        &- \kappa\langle n_j^4 \rangle
        + \kappa^2 \Big\rbrack - (m_4)^2.
\end{align}

Using the Gaussian noise identities reduces the expression to,
\begin{align}
    V_{\hat{m}_4} = &\frac{1}{N_{pix}^2}
        \sum_{i=0}^{N_{pix}}\sum_{j=0}^{N_{pix}}
        \Big\lbrack\mu_i^4\mu_j^4 
        + 6\mu_i^4\mu_j^2\sigma_j^2 
        + 3\mu_i^4\sigma_j^4 
        - \kappa\mu_i^4 \nonumber\\
        &+ 16\mu_i^3\mu_j^3\delta_{ij}\sigma_j^2 
        + 48\mu_i^3\mu_j\delta_{ij}\sigma_i^4
        + 6\mu_i^2\mu_j^4\sigma_i^2 \nonumber\\ 
        &+ 36\mu_i^2\mu_j^2(1+2\delta_{ij})\sigma_i^2\sigma_j^2 
        + 6\mu_i^2(3+12\delta_{ij})\sigma_i^2\sigma_j^4
        \nonumber\\
        &- 6\kappa\mu_i^2\sigma_i^2 
        + 48\mu_i\mu_j^3\delta_{ij}\sigma_j^4
        + 240\mu_i\mu_j\delta_{ij}\sigma_i^6 \nonumber\\ 
        &+ 3\mu_j^4\sigma_i^4 
        + 6\mu_j^2(3+12\delta_{ij})\sigma_i^4\sigma_j^2 
        + (9+96\delta_{ij})\sigma_i^4\sigma_j^4 \nonumber\\
        &- 3\kappa\sigma_i^4
        - \kappa\mu_j^4
        - 6\kappa\mu_j^2\sigma_j^2
        - 3\kappa\sigma_j^4 
        + \kappa^2 \Big\rbrack \nonumber \\
        &- (m_4)^2.
\end{align}

Doing the summation to perform index conversion via $\delta_{ij}$, substituting all $\frac{1}{N_{pix}}\sum_{i=0}^{N_{pix}}\mu_i^k$ terms with the p-th moments $m_p$ and $\sigma_i$ with $\sigma_{noise}$, re-substituting $\kappa = 3N_{pix}V_{\hat{m_2}}/2 = 6 m_2 \sigma_{noise}^2 + 3 \sigma_{noise}^4$, and cancelling out many terms will yield the estimator variance of the 4th moment,
\begin{equation}
\begin{split}
    V_{\hat{m}_4} = \frac{8}{N_{pix}}
    (2 m_6 \sigma_{noise}^2 + 21 m_4 \sigma_{noise}^4 \\
    + 48 m_2 \sigma_{noise}^6 + 12 \sigma_{noise}^8).
\end{split}
\end{equation}

The estimator covariance between 2nd and 4th moment can be found in a similar manner, resulting in,
\begin{equation}
    C_{\hat{m}_2\hat{m}_4} = \frac{4}{N_{pix}}
        (2 m_4 \sigma_{noise}^2 + 9 m_2 \sigma_{noise}^4
        + 3 \sigma_{noise}^6).
\end{equation}

The estimator variance of the moments can then be propagated to the kurtosis with Equation~\ref{eq:taylor} to obtain,
\begin{equation}
    V_{S_4} = \frac{1}{(m_2)^4} V_{\hat{m}_4}
    + 4 \frac{(m_4)^2}{(m_2)^6} V_{\hat{m}_4}
    - 4 \frac{m_4}{(m_2)^5} C_{\hat{m}_2\hat{m}_4}.
\end{equation}

\begin{table}
\centering
\begin{tabular}{l l}
    \hline
    $\begin{array}{l l l l}
        \langle n_i n_j^4 \rangle 
            && \langle n_i^5 \rangle=0 & (i=j) \\
            && \langle n_i \rangle\langle n_j^4 \rangle=0 & (i \neq j)
    \end{array}$ & $\bigg\rbrace0$ \\
    $\begin{array}{l l l l}
        \langle n_i^2 n_j^4 \rangle 
            && \langle n_i^6 \rangle=15\sigma_i^6 & (i=j) \\
            && \langle n_i^2 \rangle\langle n_j^4 \rangle=3\sigma_i^2\sigma_j^4 & (i \neq j)
    \end{array}$ & $\bigg\rbrace(3+12\delta_{ij})\sigma_i^2\sigma_j^4$ \\
    $\begin{array}{l l l l}
        \langle n_i^3 n_j^4 \rangle 
            && \langle n_i^7 \rangle=0 & (i=j) \\
            && \langle n_i^3 \rangle\langle n_j^4 \rangle=0 & (i \neq j)
    \end{array}$ & $\bigg\rbrace0$ \\
    $\begin{array}{l l l l}
        \langle n_i^4 n_j^4 \rangle 
            && \langle n_i^8 \rangle=105\sigma_i^8 & (i=j) \\
            && \langle n_i^4 \rangle\langle n_j^4 \rangle=9\sigma_i^4\sigma_j^4 & (i \neq j)
    \end{array}$ & $\bigg\rbrace(9+96\delta_{ij})\sigma_i^4\sigma_j^4$ \\
    \hline
\end{tabular}
\caption{\label{tab:noise}Additional Gaussian noise identities for derivation of estimator variance of kurtosis. Please see Table A1 in \cite{Watkinson:2014jv} for more identities.}
\end{table}

\bibliography{references}

\end{document}